\documentclass[
  aps,
  prx,
  twocolumn,
  preprintnumbers,
  longbibliography,
  superscriptaddress,
  amsmath,
  amssymb,
]{revtex4-2}

\usepackage[%
  colorlinks=true,
  urlcolor=blue,
  linkcolor=magenta,
  citecolor=blue
]{hyperref}

\usepackage{physics}
\usepackage{tikz}
\usepackage{ifthen}
\usetikzlibrary{positioning}
\usetikzlibrary{decorations.markings}

\usepackage{xfrac}
\usepackage{braket}
\usepackage{amsmath}
\usepackage{amsthm, amssymb}
\usepackage{graphicx}
\usepackage{bm}
\usepackage{array}
\usepackage{subcaption}
\usepackage{bbm}
\usepackage[normalem]{ulem}
\newcommand{\TNS}{\text{TNS}}
\newcommand{\virtual}{\text{virtual}}
\renewcommand{\vec}[1]{\bm{\mathbf{#1}}}

\newcommand{\MM}[4]{\begin{bmatrix}#1 & #2 \\ #3 & #4\end{bmatrix}}

\renewcommand{\sp}{\hspace{5mm}}
\newcommand{\A}{\mathcal{A}}
\newcommand{\DMRGsq}{$\text{DMRG}^2$ }
\newcommand{\B}{\mathcal{B}}
\newcommand{\D}{\mathcal{D}}
\newcommand{\inn}{\text{in}}
\newcommand{\out}{\text{out}}
\newcommand{\posdiag}{\scalebox{0.8}{$\diagup$}}
\newcommand{\negdiag}{\scalebox{0.8}{$\diagdown$}}

\tikzset{midarrow/.style={
    decoration={markings,
        mark= at position 0.625 with {\arrow{#1}} ,
    },
    postaction={decorate}
  }
}
\begin{document}

\title{Alternating and Gaussian fermionic Isometric Tensor Network States}

\author{Yantao Wu}
\email{yantaow@iphy.ac.cn}
\affiliation{%
Institute of Physics, Chinese Academy of Sciences, Beijing 100190, China
}%

\author{Zhehao Dai}
\affiliation{%
Department of Physics, University of California, Berkeley, CA 94720, USA
}%
\affiliation{%
Department of Physics and Astronomy, University of Pittsburgh, PA 15213, USA}%
\author{Sajant Anand}
\affiliation{%
Department of Physics, University of California, Berkeley, CA 94720, USA
}%

\author{Sheng-Hsuan Lin}
\affiliation{%
Department of Physics, TFK, Technische Universit{\"a}t M{\"u}nchen,
James-Franck-Stra{\ss}e 1, 85748 Garching, Germany
}%
\author{Qi Yang}
\affiliation{%
Institute of Physics, Chinese Academy of Sciences, Beijing 100190, China
}%
\affiliation{%
University of Chinese Academy of Sciences, Beijing 100049, China
}%
\author{Lei Wang}
\affiliation{%
Institute of Physics, Chinese Academy of Sciences, Beijing 100190, China
}%

\author{Frank Pollmann}
\affiliation{%
Department of Physics, TFK, Technische Universit{\"a}t M{\"u}nchen,
James-Franck-Stra{\ss}e 1, 85748 Garching, Germany
}%

\author{Michael P. Zaletel}
\affiliation{%
Department of Physics, University of California, Berkeley, CA 94720, USA
}%
\affiliation{Material Science Division, Lawrence Berkeley National Laboratory, Berkeley, CA 94720, USA}

\date{\today}

\begin{abstract}
  Isometric tensor networks in two dimensions enable efficient and accurate study of quantum many-body states, yet the effect of the isometric restriction on the represented quantum states is not fully understood. 
  We address this question in two main contributions.
  First, we introduce an improved variant of isometric tensor network states (isoTNS) in two dimensions, where the isometric arrows on the columns of the network alternate between pointing upward and downward, hence the name {\it alternating isometric tensor network states}.
  Second, we introduce a numerical tool -- {\it isometric Gaussian fermionic TNS} (isoGfTNS) -- that incorporates isometric constraints into the framework of Gaussian fermionic tensor network states.
  We demonstrate in numerous ways that alternating isoTNS represent many-body ground states of two-dimensional quantum systems significantly better than the original isoTNS.
  First, we show that the entanglement in an isoTNS is mediated along the isometric arrows and that alternating isoTNS mediate entanglement more efficiently than conventional isoTNS.
  Second, alternating isoTNS correspond to a deeper, thus more representative, sequential circuit construction of depth $\mathcal{O}(L_x \cdot L_y)$ compared to the original isoTNS of depth $\mathcal{O}(L_x + L_y)$.
  Third, using the Gaussian framework and gradient-based energy minimization, we provide numerical evidences of better bond-dimension scaling and variational energy of alternating isoGfTNS for ground states of various free fermionic models, including the Fermi surface and the band insulator.
  Finally, benchmarking on the transverse field Ising model, we demonstrate that alternating isoTNS provides substantially improved performance and stability relative to the original isoTNS for ground state search algorithm in interacting systems.
\end{abstract}

\maketitle


\section{Introduction}

\begin{figure}[bt]
  \begin{subfigure}{0.25\textwidth}
  \includegraphics[scale=0.9]{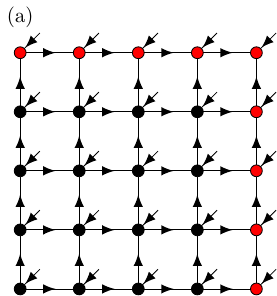}
  \captionlistentry{}
  \label{fig:uni-isotns}
  \end{subfigure}%
  \begin{subfigure}{0.25\textwidth}
  \includegraphics[scale=0.9]{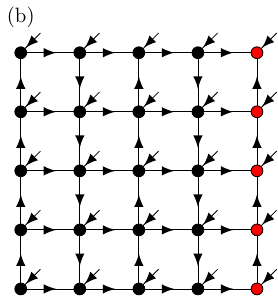}
  \captionlistentry{}
  \label{fig:alt-isotns}
  \end{subfigure}
  \begin{subfigure}{0.5\textwidth}
  \vspace*{0.25cm}
  \includegraphics[scale=1.]{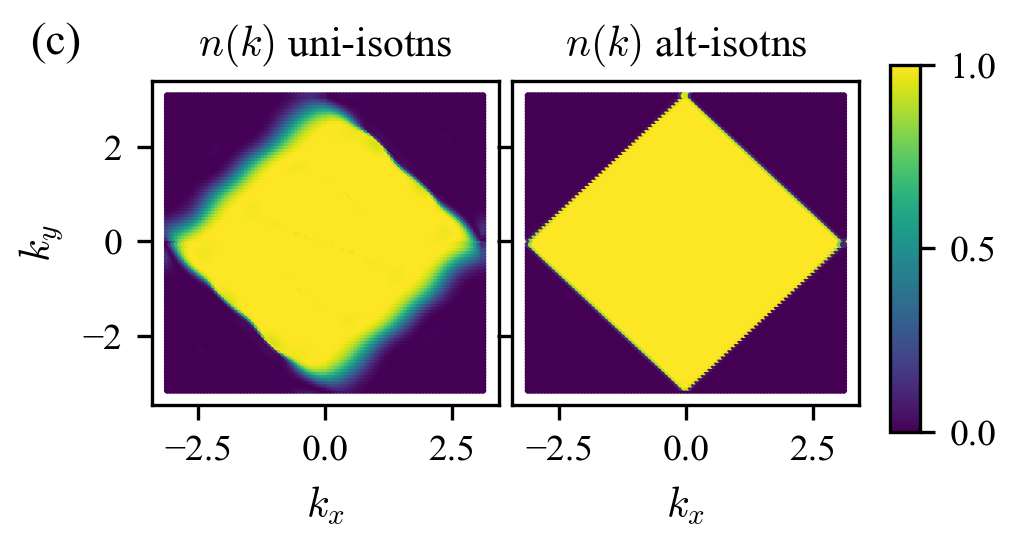}
  \captionlistentry{}
  \label{fig:FermiSurface}
  \end{subfigure}
  \caption{A $5\times 5$ uniform (a) and alternating (b) isoTNS with open boundary condition. The orthogonality center is the single, upper right tensor with all incoming legs. {The orthogonality hypersurface is the set of red tensors.} Note for the alt-isoTNS that the isometry arrow {direction} of the columns alternates, while for uni-isoTNS all the column arrows point up.
  (c) Contour plot of the $k$-space occupation number of the uni- (left) and alt-isoTNS (right) at bond dimension $\chi=16$, representing the Fermi surface ground state of Eq.~\ref{eq:fermi_surface}.
  }
\label{fig:isotns}
\end{figure}

Recently, two-dimensional (2D) tensor network states (TNS)~\cite{verstraete2004renormalization,niggemann1997quantum,nishino1998density,sierra1998density} have become a valuable tool for studying strongly interacting quantum phases of matter~\cite{ran2020tensor,cirac2021matrix,banuls2023tensor,2308.12358}.
Despite the usefulness of 2D TNS, several outstanding research challenges remain that limit their applicability, namely reducing the complexity of contracting the tensor network to compute physical observables and improving the convergence speed in optimizing the variational tensors.
For one-dimensional (1D) TNS, i.e., matrix product states (MPS), the isometric form is the key to the stability and efficiency of ground state MPS algorithms such as the density matrix renormalization group (DMRG)~\cite{white1992density, DMRG_review} and time evolution MPS algorithms based on the time-evolving block-decimation~\cite{TEBD} or the time-dependent variational principle~\cite{TDVP}. 
Thus, it is natural to consider the generalization of isometric form to 2D. 
The most straightforward version was proposed in~\cite{zaletel2020isometric} shown in Fig.~\ref{fig:uni-isotns}, known as isometric tensor network states (isoTNS). 
Similar versions are also proposed in~\cite{isoTNS_Garnet, isoTNS_Miles}.

In an isoTNS, each bond in the 2D network is assigned an arrow, and each tensor, when viewed as a map $A$ from the outgoing legs to the incoming legs, is required to be an isometric linear map, $A^\dagger A = \mathbbm{I}$.
In particular, all physical legs are incoming.
This allows both the efficient {\it exact} contraction of the norm of the 2D TNS and evaluation of expectation values of local operators on the orthogonal hypersurface~\cite{zaletel2020isometric}.
In 1D, every MPS can be exactly brought into the isometric form with the same maximal bond dimension through gauge transformations. 
In other words, the manifolds of generic MPS of bond dimension $\chi$ and MPS in isometric form of bond dimension $\chi$ are the same.
In 2D, the isoTNS is a more restricted ansatz than general TNS with the same bond dimension, and the generic and isometric manifolds of dimension $\chi$ are believed to be inequivalent.
It is an open question on how the representation power of TNS is affected by imposing the isometric constraint.

Quantum computation provides a separate motivation for studying the representability of isoTNS.
Any 1D MPS of length $L$ can be prepared on a quantum computer by a local sequential circuit of depth $L$\cite{schon2005sequential}. 
This is not true for 2D TNS unless the network has an isometric form.
Thus, it is prudent to understand the impact isometric constraints can have on the resulting sequential circuits, especially as the representability of sequential quantum circuits has recently attracted significant interests~\cite{liu2023topological,chen2024sequential_gap,chen2024sequential_chiral}.

Moreover, there actually exists freedom in choosing the isometric direction of the tensors while maintaining 
the algorithmic advantages and sequential-circuit representations.
For examples, see Fig.~\ref{fig:uni-isotns} and Fig.~\ref{fig:alt-isotns} where we show the networks of primary interest in this work.
This network design choice has not been explored in previous isoTNS works~\cite{zaletel2020isometric,lin2021efficient,iMM,isofTNS,kadow2023isometric}.
However, this turns out \textit{not} to be a gauge degree of freedom and can significantly affect the representability of the TNS ansatz and sequential circuit.
To understand the effect of different isometric forms, we provide insight from two perspectives -- entanglement structure and the corresponding sequential quantum circuits -- and support this with numerical evidence of representing ground states of both non-interacting fermionic and interacting spin systems.

In this paper, we show that in an isoTNS, entanglement is, in a sense made precise later, mediated \textit{along} the isometric arrows. 
An isoTNS is capable of representing almost any short-ranged correlated state, but it can be inefficient in capturing entanglement \textit{perpendicular} to its isometric directions.
This severely limits the application of isoTNS to some of the most important physical systems like Fermi liquids.
Here, we propose a novel isometric ansatz, alternating isoTNS (alt-isoTNS), shown in Fig.~\ref{fig:alt-isotns}, which greatly improves the representability of isoTNS while keeping all algorithmic advantages over general TNS.
To differentiate from alt-isoTNS, hereafter, we call the original isoTNS~\cite{zaletel2020isometric}, shown in Fig.~\ref{fig:uni-isotns}, the uniform isoTNS (uni-isoTNS). 

{Another independent} contribution of this paper is to introduce a tool to efficiently study the representability of isoTNS --- isometric Gaussian fermionic tensor network states (isoGfTNS). 
{IsoGfTNS is a variational ansatz resulting from imposing isometric constraint on the Gaussian fermionic tensor network states (GfTNS).}
The general problem of 2D TNS representability can be very difficult to study due to high computational complexity in optimizing a generic, {interacting} 2D TNS.   
{However, analysis can be performed by imposing the Gaussian constraint on the variational ansatz, which simplifies the calculation, and by choosing free fermion systems as a benchmark problem, where the exact solution exists~\cite{fTNS}.}
Since the key quantity restricting the TNS representability is the entanglement {}, free fermions, which can be highly entangled while still exactly solvable, are good non-trivial {benchmark problems} to test TNS representability in a controlled setting. 
In fact, significant understanding of TNS representability has been gained within the GfTNS formalism, with recent studies focusing on chiral states~\cite{GfTNS_chiral, Dubail_Read}, {$d$-wave superconductivity~\cite{GfTNS_dwave}, and} the Fermi surface~\cite{TN_Fermi_Surface}.
{Related discussion on frustration-free free fermions has also recently appeared \cite{free_free}.}

In this paper, we explain how the isometric property can be enforced as a additional constraint on top of Gaussianity within the GfTNS formalism and provide a gradient-based numerical algorithm for finding the ground state of free fermion Hamiltonians with isoGfTNS.
We use this global optimization algorithm to assess the representability of alt-isoTNS versus uni-isoTNS, free from the pitfalls of local optimization methods used for interacting quantum systems.
Using the isoGfTNS algorithm, we show that alt-isoTNS clearly captures the Fermi surface of a free-fermion metal on square lattice significantly better than uni-isoTNS.
As seen in Fig.~\ref{fig:FermiSurface}, the Fermi surface found by alt-isoTNS is demonstrably sharper and more symmetric.

{Despite the focus of this work on developing new classical numerical methods, we would like to highlight its origin and impact on quantum information. Due to the equivalence of isoTNS and sequential quantum circuits~\cite{wei2022sequential,chen2024sequential_gap}, the two main contributions of the paper have the following implications on the study of sequential quantum circuits:
(i) Gaussian fermion isoTNS: The newly proposed isoGfTNS framework provides an efficient tool to study the representability of sequential quantum circuits for many-body quantum systems, particularly the effect of circuit geometry on the circuit representability.
(ii) Alt-isoTNS: The better representability of alt-isoTNS over uni-isoTNS indicates that the resulting sequential quantum circuit layout outperforms the original sequential quantum circuit. Given a fixed number of gates, the new geometry improves the accuracy of low-energy quantum state preparation, which is the crucial building block for many quantum applications, such as the determination of spectral functions through quenches and ground-state energies via quantum phase estimation.
In fact, the correspondence to sequential quantum circuits originally motivated this work. Through the lens of quantum circuits, we ask the question, what happens if we construct an isoTNS ansatz with different quantum circuit depth scaling?
This, then, leads to our new ansatz, which significantly improves the representability of the isoTNS ansatz for strongly entangled many-body ground states, cf. Fig. \ref{fig:FermiSurface} and the stability of the classical ground-state algorithm, cf. Fig. \ref{fig:DMRGsweepdependence}.
}

The ability of isoTNS to be efficiently optimized classically and prepared on quantum devices makes it a strong candidate for quantum tasks that require efficiently obtainable classical inputs.  
Quantum quenches are one such example, where the ground state is first obtained classically, and then prepared on a quantum computer for subsequent dynamics -- a task quantum computers excel.  
In this context, the isoGfTNS is also useful beyond free (Gaussian-fermionic) systems, particularly when the subsequent dynamics is interacting (non-Gaussian).

This paper is organized as follows.
In Sec.~\ref{sec:altisotns}, we introduce the alt-isoTNS and explain how it mediates entanglement differently than uni-isoTNS. 
We then discuss the difference between alt-isoTNS and uni-isoTNS in terms of sequential quantum circuits and holographic state preparation.
We next demonstrate that all quantum double fixed points have exact alt-isoTNS representations.
The section ends with the discussion on how the usual uni-isoTNS algorithms for interacting quantum systems can be easily modified for the alt-isoTNS ansatz.
In Sec.~\ref{sec:isogftns}, we introduce the isoGfTNS framework and the necessary optimization algorithms.   
This section {is} of independent interest from alt-isoTNS, as
{
we can consider arbitrary, consistent isomeric conditions and explore the representability of different isoTNS within this framework.} 
In Sec.~\ref{sec:results}, we first study non-interacting models, including the 2D Fermi surface and gapped free fermion insulators, providing numerical evidence of the superior representation of alt-isoGfTNS compared to uni-isoGfTNS.
We next perform a study of the interacting 2D transverse field Ising model where alt-isoTNS demonstrates significantly better performance {and algorithmic stability} than uni-isoTNS in finding the ground state.
Finally, in Sec.~\ref{sec:discussion}, we discuss future directions for isometric tensor networks and conclude.

\section{Alternating isometric tensor network states
\label{sec:altisotns}}

Isometric tensor network states are tensor network states where each tensor satisfies certain isometric constraints. 
In 1D, there are two natural choices for isometric constraints, often referred to as left-canonical form and right-canonical form.
A tensor $A_{p,l,r}$ is called left-canonical if $\sum_{p,l} A_{p,l,r}A^*_{p,l,r'} = \delta_{r,r'}$, where $p, l, r$ are the physical, left, and right index of the tensor.
Correspondingly, a right canonical tensor satisfies $\sum_{p,r} A_{p,l,r}A^*_{p,l',r} = \delta_{l,l'}$.
It is convenient to represent this isometric condition graphically in terms of the arrows.
The arrows indicate how legs of the tensors, i.e.  indices, should be grouped so that the resulting matrix is an isometry; in our convention, the matrix $A$ with all incoming and outgoing legs forming the row and column indices, respectively, is isometric; $A^\dagger A = \mathbbm{I}$.
Below, we refer to the arrow direction as the isometric direction.

In 1D, any matrix product state can be written in a left-canonical (or right-canonical) form with the same bond dimension.
In 2D, there is a non-trivial freedom of choice in the isometric directions of the tensors that has been unexplored up until now.
Unlike in the 1D case, 2D isoTNS form a strict subset of 2D TNS, and different choices of isometric directions give different subsets of general TNS.
In a conventional uni-isoTNS, the isometric arrows of each isometry (the black tensors in Fig. \ref{fig:uni-isotns}) all point towards the same direction,~e.g.,~upper right in Fig.~\ref{fig:uni-isotns}.
In an alt-isoTNS, we keep the horizontal isometric arrows pointing rightward but alternate the vertical isometric arrows between upward and downward pointing on even and odd columns.
{With open boundary condition, there exists an orthogonality hypersurface (OHS) of tensors, which only has arrows point inward, depicted in red in Fig.~\ref{fig:isotns}.
The OHS is composed of a column and a row in an uni-isoTNS, while it is only composed of a column in an alt-isoTNS.
The OHS can be treated as an MPS embedded in the 2D network, and the orthogonality center (OC) is the single tensor with all incoming legs.}
As will be shown, alt-isoTNS retains all the algorithmic efficiencies that enable efficient classical manipulation and optimization of uni-isoTNS but greatly improves wave function representability.

\begin{figure}[tb]
  \begin{subfigure}{0.25\textwidth}
  \caption{uni-isoTNS}
  \includegraphics[scale=0.88]{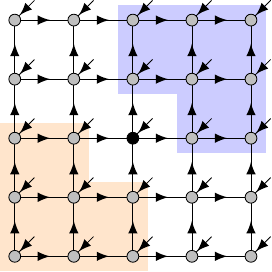} 
  \label{fig:causal-uni}
  \end{subfigure}%
  \begin{subfigure}{0.25\textwidth}
    \caption{alt-isoTNS}
  \includegraphics[scale=0.88]{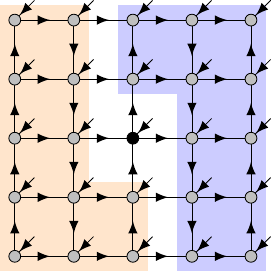}
  \label{fig:causal}
  \end{subfigure}
\caption{Causal structure of uni- (a) and alt-isoTNS (b). For the single site tensor colored in black, its future (past) light cone is colored on orange (blue). The uncolored tensors belong to the space-like region outside the light cone.
}
\label{fig:causal_cone}
\end{figure}

A first hint of the physical difference resulting from the isometric structure is in the causal light cones shown in Fig.~\ref{fig:causal_cone}.
Here we define the \emph{future light cone} {of a tensor} as the sites whose on-site expectation values of physical observables are affected by the change of the tensor (shown in black).
The future light cone, highlighted in orange, of a tensor includes all tensors that can be reached by going \textit{against} the flow of arrows on the legs.
The \emph{past light cone}, highlighted in blue, are the set of tensors that can affect the on-site expectation values of physical observables of the current (black) tensor, reached by going \textit{with} the flow of arrows.
{A tensor is \emph{space-like} separated from the black tensor if it is neither in the causal past or future of the black tensor.}
 {In alt-isoTNS no tensors are space-like separated from the black tensor, while in uni-isoTNS there exist two regions of space-like separation from the black tensor.}
This light cone structure follows from the isometric structure and have a direct impact on both the entanglement structure and the sequential quantum circuit construction.

\subsection{Mediating entanglement along diagonals}
\label{sec:diagonal}

We next discuss the marked differences in entanglement structure between uni-isoTNS and alt-isoTNS, focusing on the isometric structures shown in Fig.~\ref{fig:uni-isotns} and Fig.~\ref{fig:alt-isotns}.
A uni-isoTNS  prefers to entangle physical sites along the $(1,1)$ diagonal, which aligns with its light cone.
An alt-isoTNS, on the other hand, does not have this anisotropy.
To make this anisotropy manifestly clear, consider a concrete, albeit manufactured, example state $\ket{\posdiag}$, which is an $L_x \times L_y$ 2D state formed from the tensor product of $L_x + L_y$ decoupled chains along $(1,1)$-diagonals, and contrast it with the state $\ket{\negdiag}$, the analogous state formed by a tensor product of $L_x + L_y$ decoupled chains along $(1,-1)$ direction. 
For simplicity, we assume each decoupled chain is an 1D area-law state and thus can be efficiently represented by an MPS with a finite bond dimension $\chi$.

\begin{figure}[tb]
  \begin{subfigure}{0.25\textwidth}
  \caption{uni-isoTNS, $\ket{\posdiag}$ state}
  \includegraphics[scale=1.0]{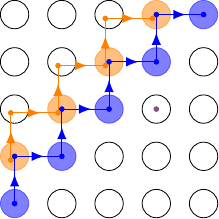}
  \label{fig:uni-mps-posdiag}
  \end{subfigure}%
  \begin{subfigure}{0.25\textwidth}
  \caption{alt-isoTNS, $\ket{\posdiag}$ state}
  \includegraphics[scale=1.0]{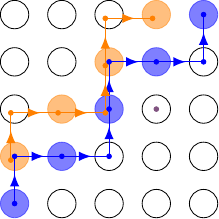}
  \label{fig:alt-mps-posdiag}
  \end{subfigure} 
  \begin{subfigure}{0.25\textwidth}
  \caption{uni-isoTNS, $\ket{\negdiag}$ state}
  \includegraphics[scale=1.0]{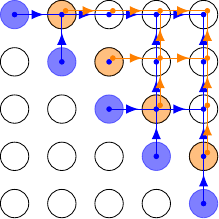}
  \label{fig:uni-mps-negdiag}
  \end{subfigure}%
  \begin{subfigure}{0.25\textwidth}
   \caption{alt-isoTNS, $\ket{\negdiag}$ state}
  \includegraphics[scale=1.0]{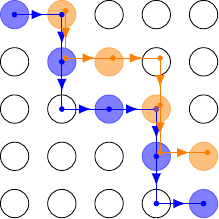}
  \label{fig:alt-mps-negdiag}
  \end{subfigure}
\caption{Uni (a,c) and alt-isoTNS (b,d) representions of decoupled chains along the (1,1)-diagonals (a,b) and along the (1,-1)-diagonals (c,d). Two chains, one in blue and one in orange, are shown.
The physical indices are suppressed for visual clarity. Crucially, the uni-isoTNS representation of $\ket{\negdiag}$ requires a MERA structure, while the alt-isoTNS representation mediates all entanglement locally.}
\label{fig:diagonal-alt}
\end{figure}

Since both $\ket{\posdiag}$ and $\ket{\negdiag}$ are tensor products of 1D states, we can utilize the MPS representation of the 1D states to construct the 2D isoTNS representation.
For an uni-isoTNS, we can represent $\ket{\posdiag}$ efficiently by embedding the MPS representation of each chain directly into the 2D state. We show the MPS representation in isometric form for two (1,1)-chains, e.g., the blue and the orange MPS, in Fig.~\ref{fig:uni-mps-posdiag}.
Such representations can be efficiently embedded as part of a uni-isoTNS while remaining consistent with isometric (arrow) direction. 
A different construction for $\ket{\posdiag}$ is given for alt-isoTNS in Fig.~\ref{fig:alt-mps-posdiag}.
Due to the isometric constraint on the columns, we can only pass the information upward at specific columns.
As a result, uni-isoTNS is slightly more efficient than alt-isoTNS in mediating the entanglement along the $(1,1)$ direction:
uni-isoTNS requires bond dimension $\chi$ for each bond while alt-isoTNS requires bond dimension $\chi^2$ when two MPS embeddings overlap.

A drastic difference shows up when we consider the case of the $\ket{\negdiag}$ state.
For an alt-isoTNS, we can again represent $\ket{\negdiag}$ efficiently by embedding the MPS representation of each chain directly into the 2D state.
We show the MPS representation in isometric form for (1,-1)-chains, e.g., the blue and the orange MPS in Fig.~\ref{fig:alt-mps-negdiag}.
Similar to the construction for $\ket{\posdiag}$, we can only pass the information downward at specific columns, and therefore we require a bond dimension $=\chi^2$, independent of system size $L$.
However, the physical sites along a coupled $(1,-1)$ diagonal chain are space-like separated in uni-isoTNS.
Due to the lack of downward pointing isometry arrows, the only way to encode non-trivial correlations between two sites is to encode the information in a tensor in the common causal past.
This requirement is particularly bad for the \textit{exact} representation of decoupled chains on an \textit{system with open boundary condition}.
For the 5-site (blue) and 4-site (orange) chains shown in Fig.~\ref{fig:uni-mps-negdiag}, only the single tensor in the top right corner can mediate the entanglement between the two ends of the chains.
For an $L\times L$ system with open boundary condition, we need at least a bond dimension $2^{L-1}$ at the top-right corner to exactly represent the $\ket{\negdiag}$ state.

The requirement of the enormous bond dimension at the corner for uni-isoTNS can be significantly alleviated if we consider a periodic tensor network or allow minor errors in the representation.
To better compare the average bond dimension of approximate uni-isoTNS and alt-isoTNS representations, we consider the case where each decoupled chain in the $\ket{\negdiag}$ state is a near-critical state with a long correlation length $\xi$; the nearby critical point is assumed to be described by a conformal field theory with a central charge $c$.
According to Ref.~\cite{entanglement_scaling}, the MPS representing this state has a bond dimension $\chi\sim \xi^{1/\kappa}$, where $\kappa$ is a function of the central charge.
Similar to Fig~\ref{fig:alt-mps-negdiag}, each diagonal chain uses $2L$ bonds on a $L\times L$ periodic alt-isoTNS; The product of all bond dimensions for a single chain is $\chi^{2L}$.
Stacking all chains, the \textit{geometric average} of the $2L^2$ bond dimensions is $\chi_\text{alt} = \chi^{2L\times L/2L^2} = \chi$.
For uni-isoTNS, an efficient way to represent the diagonal chain while respecting the isometric structure is to embed a MERA into the isoTNS. As shown in Fig.~\ref{fig:uni-mps-negdiag}, the disentangling direction of the MERA corresponds to the $(1,1)$ direction of the isoTNS; they have consistent isometric arrows. We estimate the bond dimension of the MERA to be {$\exp(c/6)$} \cite{holography_cft_bond_dimension} and the depth of the MERA to be $O(\ln(\xi))$. At level $d$ of the MERA, we have $L/2^d$ sites left, but the distance between neighbors becomes $2^d$; therefore, the embedding uses $2L$ bonds at each level. This counting leads to a geometric average of bond dimensions $\chi_\text{uni} \sim e^{\frac{c}{6}O(\ln(\xi))} \sim \chi_\text{alt}^\eta$.
Determining the exact value of $\eta$ is difficult due to $O(1)$ uncertainties in the estimation.
Since we can collapse each MERA into an MPS without changing the product of the bond dimensions, $\eta$ must be greater or equal to 1.
In Sec~\ref{sec:isogftns}, we introduce Gaussian fermion isoTNS{ and apply it} in Sec.~\ref{sec:free_fermion_hopping_1_1} to numerically test the argument presented here. Our numerical results suggest $\eta> 1$.
Unlike the naive argument, $\eta$ is independent of system size.

In conclusion, alt-isoTNS behaves the same for decoupled chains in the $(1,1)$ direction and decoupled chains in the $(1,-1)$ direction. We expect uni-isoTNS to have an advantage over alt-isoTNS for mediating entanglement in the $(1,1)$ direction but a disadvantage for the $(1,-1)$ direction.
For an isotropic system, there are entangled degrees of freedom with arbitrary separations, and we expect the dominant error to come from the worst direction.
Since alt-isoTNS is more symmetric in mediating entanglement, we expect it to be an overall better ansatz.
In Sec.~\ref{sec:DMRG}, we show numerical evidence supporting this argument for interacting, isotropic spin models.

\subsection{Sequential quantum circuits \label{subsec:SQC}}
\begin{figure}[tb]
  \begin{subfigure}{0.25\textwidth}
  \caption{right-canonical MPS}
  \vspace{6mm}
  \includegraphics[scale=1.2]{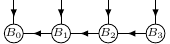} 
  \end{subfigure}%
  \begin{subfigure}{0.25\textwidth}
  \caption{sequential circuit}
  \includegraphics[scale=1.2]{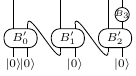}
  \end{subfigure}
\caption{The sequential circuit representation of an isometric MPS. $B_i'$ is a unitary matrix obtained from adding orthogonal columns to the isometric matrix $B_i$.
}
\label{fig:sqc_mps}
\end{figure}
\begin{figure}[t]
  \includegraphics[scale=1.1]{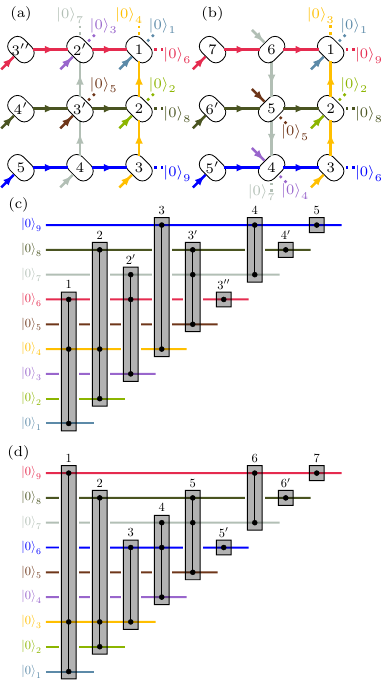}
\caption{(a) Uni- and (b) alt-isoTNS when viewed as sequential quantum circuit to prepare a 9 qubit state. All bond dimensions are $\chi=2$, so each line represents a qubit.
Different qubits are drawn in different colors.
The label of the tensors corresponds to the time-step when the gate can be applied.
Dotted lines indicate input qubits in the state $\ket{0}$.
Due to isometry conditions on each tensor, the vertical bond dimension in the left-most column is $\chi=1$.
(c) [(d)] Quantum circuit diagrams for uni-[alt-]isoTNS, where isometric tensors have been expanded into unitary gates acting on up to 3 qubits by adding orthogonal columns.
Gates act on the qubits that continue through the rectangle. Gates with the same number act on disjoint sets of qubits and thus can be applied simultaneously.
}
\label{fig:circuitl}
\end{figure}

Beyond their utility for numerical and analytical studies, isometric tensor networks have a direct correspondence to sequential quantum circuits and provide concrete unitary circuits for qubit-efficient state preparation on quantum computers~\cite{schon2005sequential,schon2007sequential,banuls2008sequentially,wei2021generation,wei2022sequential}.
This has been demonstrated experimentally in 1D for both MPS~\cite{foss-feig2021} and entanglement renormalization tensor networks~\cite{anand2023,haghshenas2023} and explored numerically in 2D~\cite{slattery2021}.

To recap the general principle behind mapping isometric tensor networks to quantum circuits, consider an MPS in right-canonical form (left pointing isometry arrows) with bond dimension $\chi$ and physical Hilbert space dimension $d=2$, as shown in Fig. \ref{fig:sqc_mps}. 
Starting from the left end, the first isometric tensor can be expanded into a unitary acting on a physical qubit initialized to some fixed state, typically $\ket{0} \in \mathbb{C}^{2}$ and a bond qudit of dimension $\chi$, typically also initialized to $\ket{0} \in \mathbb{C}^{\chi}$;
the bond qudit can also be viewed as a register of $\log_2(\chi)$ qubits. 
The isometry is expanded into a unitary by adding orthogonal columns that dictate the action of the gate when the physical qubit is in any state orthogonal to the desired one, e.g. $\ket{1}$.
The unitary derived from the tensor on the second site acts on a fresh physical qubit in $\ket{0}$ and the bond qubit from the first site, which carries correlations from site to site. 
In this way, a length $L$ MPS can be sequentially prepared by a depth $L$ circuit, nominally using $L$ qubits~\cite{schon2005sequential}.

This idea directly carries over to all isometric tensor networks and in particular to the 2D geometries depicted in Fig.~\ref{fig:isotns}. 
A bond dimension $\chi$ uni-isoTNS corresponds to a depth $\mathcal{O}(L_x + L_y)$ state preparation circuit;
for example, the $3 \times 3$ uni-isoTNS shown in Fig.~\ref{fig:circuitl}(a) corresponds to a depth 5 circuit preparing the 9 qubit state.
At each time step $t$, the unitary gates corresponding to the tensors a Manhattan distance $t$ from the {OC} (upper right) can be applied in parallel. 
Once again, the light cone picture plays an important role here.
Gates with space-like separation can be applied simultaneously.
Additionally, the notion of past and future causal cones indicates which gates in the circuit can affect physical observables on sites and thus affect the pattern of entanglement generated.
A circuit diagram for the resulting sequential quantum circuit is shown in Fig.~\ref{fig:circuitl}(c).

Alt-isoTNS corresponds to a drastically different circuit, as due to the alternating isometry conditions in the columns, typically only one unitary gate can be applied at each time step~\footnote{Due to the isometry conditions on each tensor, with physical dimension $d=2$ one cannot have a non-trivial uniform bond dimension $\chi > 1$ for all virtual legs at the boundary.
As shown in Fig.~\ref{fig:circuitl}(a) and Fig.~\ref{fig:circuitl}(b), the vertical bond dimensions of the leftmost column (i.e. furthest from the {OC}) are trivial.
Thus, tensors in the final column, when viewed as a gates of the resulting quantum circuit, quantum circuit, can be applied in parallel with other gates.
This is a boundary effect. For alt-isoTNS, only one unitary gate can be applied at each time step inside the bulk.
Therefore, the depth of an alt-isoTNS circuit is $\mathcal{O}(L_x \cdot L_y)$}.
As an example, consider the $3 \times 3$ alt-isoTNS in Fig.~\ref{fig:circuitl}(b) and the resulting sequential quantum circuit in Fig.~\ref{fig:circuitl}(d).
One sequentially acts all of the unitaries in the orthogonality column (down the rightmost column in Fig.~\ref{fig:circuitl}(b)) and then crosses over into the next column, sequentially applying gates against the direction of the isometry arrows (up the middle column in Fig.~\ref{fig:circuitl}), and so on in a snake-like pattern for the remaining columns.
Thus, alt-isoTNS corresponds to a depth $\mathcal{O}(L_x \cdot L_y)$ circuit.
This corresponds to the lack of space-like separation in alt-isoTNS, so each tensor lies in either the future or past causal cone of each other.
The quadratically deeper circuit nature of alt-isoTNS compared to uni-isoTNS could potentially lead to increased 
representability,~i.e.,~numerical performance when optimized classically, of the alternating ansatz.

The sequential quantum circuits discussed above can be combined with the holographic approach, where one prepares a $D$-dimensional quantum state using a number of qubits proportional to a $(D-1)$-dimensional cross-section of the state~\cite{foss-feig2021,kim2017}.
Roughly speaking, if all we are interested in is measuring physical expectation values and not preparing the quantum state itself for further manipulation, we can recycle qubits produced from the application of the unitary by measuring the physical qubit and reusing it for the next unitary, after reinitializing it to $\ket{0}$. 
This can drastically decrease the required number of physical qubits to actually perform quantum simulation, at the expense of requiring mid-circuit measurement and reset.
In terms of sequential circuits of MPS, expectation values can be estimated using $1 + \log_2(\chi)$ qubits, 1 for the physical qudit and the remaining to represent the bond qudit. 
For sequential circuits of uni-isoTNS and alt-isoTNS, $\mathcal{O}((L_x + L_y)\log_2(\chi)$ qubits and $\mathcal{O}(\mathrm{min}(L_x, L_y) \cdot \log_2(\chi))$ qubits are needed, respectively, for holographic simulation.
While uni-isoTNS requires more qubits, the depth of the resulting circuit is shorter, so the choice of better ansatz in practice will depend on noise levels and qubit numbers on the particular device~\cite{anand2023}.
In particular, we expect uni-isoTNS to be more noise resilient, as the smaller causal cones limit the negative effects errors in gates can have on future expectation values.

We conclude by noting that prior studies have been restricted to much smaller system sizes~\cite{du2022quantum,du2022efficient,funcke2021dimensional,wu2021expressivity,haug2021capacity}, and that sequential circuits in the literature have focused only on uniform structures~\cite{wei2022sequential,chen2024sequential}. In addition, recent theoretical works address different but complementary aspects~\cite{miao2024isometric,barthel2025absence}. By working at larger scales, introducing alternating sequential structures, and demonstrating practical optimization schemes (Sec.~\ref{sec:results}), our results extend the understanding of sequential circuits and highlight their relevance for both quantum simulation and quantum algorithm design.

\subsection{Topological phases}

We now demonstrate a simple proof that alt-isoTNS can represent a large class of topological orders in 2D. 
It has been proven that all string-net states, which contain all non-chiral topological orders with gappable edges in two dimensions~\cite{stringnet, 2D_classify}, can be exactly represented as finite bond dimension uni-isoTNS in the thermodynamic limit~\cite{stringnet_isotns}. 
This construction, however, is quite complicated.  
{
Here we present a much simpler construction for a large subset of string-net states, the quantum double states of finite group $G$~\cite{QC_anyon}, for both the uni-isoTNS and the alt-isoTNS, demonstrating the versatile representability of alt-isoTNS.} 
Up to normalization, local unitaries, and coarse-graining, the exact TNS tensor representing the ground state of the quantum double for group $G$ is found~\cite{PEPS_GS} to be   
\begin{equation}
  A_{lrdu}^{p_1 p_2 p_3 p_4}  = \sum_{g\in G} g_{l p_1} g_{u p_4} g^{-1}_{p_2 r} g^{-1}_{p_3 d},
\end{equation}
where $g$ denotes both the group element and its left regular matrix representation which is a real permutation matrix~\footnote{The left regular matrix representation of a group $G$ is defined as follows. Consider the vector space spanned by all group elements: $V = \text{span}\{g | g\in G\}$, where all $g$s are defined to be linearly independent. The matrix representation $\phi(g)$ of $g$ in $V$ is a permutation matrix whose action is defined on the basis of $V$ as $\phi(g) \ket{g'} = \ket{gg'} \forall g' \in G$. This is clearly a, possibly reducible, matrix representation of $G$, because $\phi(g) \phi(g') = \phi(gg')$. All $\phi(g)$ are traceless expect when $g = e$.}. 
The tensor legs label group elements of $G$, so the bond dimension is the order of the group, $|G|$. 
Due to coarse-graining, $p_1p_2p_3p_4$ are grouped into one physical leg on each tensor. 
If one contracts any two of the virtual legs, e.g. $l$ and $d$, and the physical leg of $A$ with that of $A^\dagger$, one gets
\begin{align}
  &\sum_{ld p_1p_2p_3p_4} A_{lrdu}^{p_1 p_2 p_3 p_4} A_{lr'du'}^{p_1 p_2 p_3 p_4} \\
  & = \sum_{h,g\in G} \Tr(g h^{-1}) (gh^{-1})_{uu'} (g h^{-1})_{rr'} \Tr(g^{-1} h) \\  
  & \propto \delta_{uu'} \delta_{rr'}
\end{align}
where we have noted that because $g^{-1}h$ is again a {matrix in the left regular representation}, it is traceless unless $g^{-1} h$ is the identity. 
This computation holds for any two virtual legs. 
The quantum double TNS is thus automatically both a uni-isoTNS and an alt-isoTNS.
Note that the quantum double tensors actually satisfy far stronger isometric constraints than either those of uni- or alt-isoTNS, as any grouping of two virtual legs and the physical legs lead to a valid isometry.

\subsection{Algorithmic modification for alternating isometric tensor network states}
\label{subsec:alt-isotns-algorithm} 

The alt-isoTNS ansatz introduced above can be implemented easily using the standard isoTNS toolbox~\cite{zaletel2020isometric, isofTNS, lin2021efficient, iMM} with minor modification.
Crucially, the computational complexity remains the same for the time evolution algorithm~\cite{zaletel2020isometric,iMM}, $\mathcal{O}(\chi^7)$ and ground state search algorithms~\cite{lin2021efficient}, $\mathcal{O}(\chi^{10})$.

There are two key subroutines in isoTNS algorithms. 
One is the splitting of the orthogonality two-column $\Lambda_{i:i+1}$, analogous to the two-site wavefunction in an MPS, into a left isometry column $A_i$ and an orthogonality column $M_{i+1}$, analogous to an isometric tensor and normalized tensor in an MPS; this is shown diagramatically in Eq.~\eqref{eq:MM}. 
$\Lambda_{i:i+1}$ carries the physical legs of columns $i$ and $i+1$, with the physical legs on column $i$ merged with $\Lambda_{i:i+1}$'s left virtual legs and those on column $i+1$ with the right virtual legs.
In the \emph{Moses Move} algorithm used to perform this splitting~\cite{zaletel2020isometric}, the resulting vertical isometric arrows in $A_i$ and $M_{i+1}$ are always reversed from the original vertical arrows in $\Lambda_{i:i+1}$:
\begin{equation}
  \newcommand{\LL}{3}      
\renewcommand{\d}{1.0}   
\renewcommand{\r}{0.1}   
\renewcommand{\a}{0.8}   
\newcommand{\dH}{0.6}    
\newcommand{\aH}{0.4}    
\begin{tikzpicture}[baseline = (X.base),every node/.style={scale=1.0},scale=1.2]
\newcommand{\x}{0}
\newcommand{\y}{0}
\draw (\LL*\d/2,\LL*\d/2) node (X) {};
\foreach \j in {0,...,\LL}
{
  \renewcommand{\x}{0}
  \renewcommand{\y}{\j*\d}
  \pgfmathsetmacro{\color}{ifthenelse(\j==0, "red", "red")}
  \draw [fill=\color] (\x,\y) circle (\r);
  \draw [midarrow={latex reversed}](\x+\r,\y) -- (\x+\r+\aH,\y); 
  \draw [midarrow={latex}](\x-\r-\aH,\y) -- (\x-\r,\y); 
  \ifthenelse{\j<\LL}{
  \draw [midarrow={latex reversed}](\x,\y+\r) -- (\x,\y+\r+\a);
  }{}
}
\draw (0,-0.5) node {\scalebox{0.8}{$\Lambda_{i:i+1}$}};
\end{tikzpicture}
\hspace{-6mm}
\approx
\hspace{6mm}
\begin{tikzpicture}[baseline = (X.base),every node/.style={scale=1.0},scale=1.2]
\draw (\LL*\d/2,\LL*\d/2) node (X) {};
\newcommand{\x}{0}
\newcommand{\y}{0}
\foreach \i in {0,...,1}
{
  \foreach \j in {0,...,\LL}
  {
    \renewcommand{\x}{\i*\dH}
    \renewcommand{\y}{\j*\d}
    \pgfmathsetmacro{\Hdir}{ifthenelse(\i == 0, "latex", "latex reversed")}
    \pgfmathsetmacro{\Vdir}{ifthenelse(\j <\LL,"latex", "latex reversed")}
    \pgfmathsetmacro{\color}{ifthenelse(\i==0, "black", "red")}
    \draw [fill=\color] (\x,\y) circle (\r);
    \draw [midarrow={\Hdir}](\x+\r,\y) -- (\x+\r+\aH,\y); 
    \ifthenelse{\j<\LL}{
    \draw [midarrow={\Vdir}](\x,\y+\r) -- (\x,\y+\r+\a);
    }{}
  }
}
\foreach \j in {0,...,\LL}
{
  \renewcommand{\x}{0*\d}
  \renewcommand{\y}{\j*\d}
  \draw [midarrow={latex reversed}](\x-\r,\y) -- (\x-\r-\aH,\y); 
}
\draw (0,-0.5) node {\scalebox{0.8}{$A_i$}};
\draw (\dH+\aH/2,-0.5) node {\scalebox{0.8}{$M_{i+1}$}};
\end{tikzpicture}

  \label{eq:MM}
\end{equation}
The vertical arrow direction in $\Lambda_{i:i+1}$ can be efficiently and exactly reversed similar to changing the isometric form of an MPS.
Thus, we can control the resulting isometric directions in $A_i$ and $M_{i+1}$ by choosing the corresponding opposite direction on $\Lambda_{i:i+1}$ as the input to the splitting algorithm.
Explicitly, when sweeping between the columns using the Moses Move, we can impose the alt-isoTNS (uni-isoTNS) structure by ensuring that $\Lambda_{i:i+1}$ has the same (opposite) vertical isometry direction as the previous isometric columns.

The second key subroutine in the isoTNS algorithm is the reverse of the splitting operator described above,~i.e.,~the merging of the orthogonality column $M_{i+1}$ and a right isometric column $B_{i+2}$ into a new orthogonality two-column $\Lambda_{i+1:i+2}$:     
\begin{equation}
  \newcommand{\LL}{3}      
\renewcommand{\d}{1.0}   
\renewcommand{\r}{0.1}   
\renewcommand{\a}{0.8}   
\newcommand{\dH}{0.6}    
\newcommand{\aH}{0.4}    
\begin{tikzpicture}[baseline = (X.base),every node/.style={scale=1.0},scale=1.2]
\draw (\LL*\d/2,\LL*\d/2) node (X) {};
\newcommand{\x}{0}
\newcommand{\y}{0}
\foreach \i in {0,...,1}
{
  \foreach \j in {0,...,\LL}
  {
    \renewcommand{\x}{\i*\dH}
    \renewcommand{\y}{\j*\d}
    \pgfmathsetmacro{\Vdir}{ifthenelse(\j <\LL,"latex", "latex reversed")}
    \pgfmathsetmacro{\color}{ifthenelse(\i==0, "red", "black")}
    \draw [fill=\color] (\x,\y) circle (\r);
    \draw [midarrow={latex reversed}](\x+\r,\y) -- (\x+\r+\aH,\y); 
    \ifthenelse{\j<\LL}{
    \draw [midarrow={\Vdir}](\x,\y+\r) -- (\x,\y+\r+\a);
    }{}
  }
}
\foreach \j in {0,...,\LL}
{
  \renewcommand{\x}{0*\d}
  \renewcommand{\y}{\j*\d}
  \draw [midarrow={latex reversed}](\x-\r,\y) -- (\x-\r-\aH,\y); 
}
\draw (0,-0.5) node {\scalebox{0.8}{$M_{i+1}$}};
\draw (\dH+\aH/2,-0.5) node {\scalebox{0.8}{$B_{i+2}$}};
\end{tikzpicture}
=
\hspace{6mm}
\begin{tikzpicture}[baseline = (X.base),every node/.style={scale=1.0},scale=1.2]
\newcommand{\x}{0}
\newcommand{\y}{0}
\draw (\LL*\d/2,\LL*\d/2) node (X) {};
\foreach \j in {0,...,\LL}
{
  \renewcommand{\x}{0}
  \renewcommand{\y}{\j*\d}
  \pgfmathsetmacro{\color}{ifthenelse(\j==0, "red", "red")}
  \draw [fill=\color] (\x,\y) circle (\r);
  \draw [midarrow={latex reversed}](\x+\r,\y) -- (\x+\r+\aH,\y); 
  \draw [midarrow={latex}](\x-\r-\aH,\y) -- (\x-\r,\y); 
  \ifthenelse{\j<\LL}{
  \draw [midarrow={latex}](\x,\y+\r) -- (\x,\y+\r+\a);
  }{}
}
\draw (0,-0.5) node {\scalebox{0.8}{$\Lambda_{i+1:i+2}$}};
\end{tikzpicture}

  \label{eq:merge}
\end{equation}
Here, for algorithmic efficiency reasons, it is advantageous if the vertical arrows on $M_{i+1}$ and $B_{i+2}$ have the same direction so that 
the resulting $\Lambda_{i+1:i+2}$ has a definite isometric form. 
If $M_{i+1}$ and $B_{i+2}$ have opposite vertical arrow directions, the direction on $M_{i+1}$ can be reversed to agree with $B_{i+2}$ before merging.  

Thus, regardless of the desired pattern of the direction of the vertical isometry arrows, it can be achieved by reversing of the isometric form on the orthogonality columns or two-columns when needed. 
This has a computational complexity $O(\eta^3 \chi^2)$, compared to the $O(\eta^3\chi^4)$ of splitting algorithm, where $\eta$ is the vertical bond dimension on the orthogonality column and $\chi$ is the bond dimension elsewhere. 
The cost of imposing a particular isometry structure is subleading, and thus alt-isoTNS is, at leading order, no more computationally expensive than uni-isoTNS.

These subroutines are used in both ground state and time evolution algorithms. 
{In particular, they are used for computing local observables in an isoTNS. 
Any operator supported on column $i$ and $i+1$ can be measured by applying MPS methods to $\Lambda_{i:i+1}$ after $\Lambda_{i:i+1}$ has been made into the OHS via the Moses Move algorithm.}

In Sec.~\ref{sec:DMRG} we investigate the performance of uni- and alt-isoTNS in representing the ground state of the 2D transverse field Ising model.

\section{Isometric Gaussian fermionic Tensor Network States}
\label{sec:isogftns}

{A separate result of this paper is the introduction of a Gaussian version of isoTNS, alternating, uniform, or otherwise.}
We will show how to incorporate the isometric constraint into the Gaussian fermionic tensor network states (GfTNS) framework, resulting in the isometric GfTNS (isoGfTNS).
{As an application, this will be used to assess the representability of alt-isoTNS versus uni-isoTNS.}
It is common to optimize translationally invariant GfTNS with variational global approach using gradient descent to study the representation power of the ansatz.
In the isoGfTNS framework, the uniform and alternating constraints can be implemented by choosing the corresponding fixed isometric form in the tensor initialization.
In the next section using these techniques, we consider various non-interacting (free-) fermion models and compare the performance of uniform and alternating isoGfTNS in representing their ground states.
This allows us to investigate the effect of isometric constraints on representability with a different angle from interacting isoTNS.

\subsection{Fermionic Gaussian states}

Let us consider free fermion system on a 2D square lattice of size $L_x \times L_y$, where each site hosts a complex fermion mode $c$. 
The information of a fermionic Gaussian state $\rho$ is entirely encapsulated in its covariance matrix: 
\begin{equation}
\Gamma^{\mu\nu}_{\vec x, \vec x'} \equiv \frac{i}{2} \Tr(\rho[\gamma^{\mu}_{\vec x}, \gamma^\nu_{\vec x'}])
\label{eq:covar}
\end{equation}
where $\gamma^{\mu}_{\vec x}$ is the (real) Majorana fermion mode on site $\vec x$:  
\begin{equation}
\gamma^{1}_{\vec x}  = c_{\vec x}^\dag + c_{\vec x}, \sp \gamma^{2}_{\vec x} = -i(c_{\vec x}^\dag - c_{\vec x})
\end{equation}
In the following discussion, we consider free fermion models with translational invariance without unit cells. 
The generalization to unit cells is straightforward, but the notation is onerous.
We discuss unit cells in Appendix \ref{subsec:unit_cell}.  
After a Fourier transform, the covariance matrix in the momentum space is: 
\begin{equation}
  \Gamma^{\mu\nu}(\vec k) \equiv \frac{i}{2} \Tr(\rho[\gamma_{\vec k}^\mu, \gamma^{\nu\dag}_{\vec k}]) = \frac{i}{2} \Tr(\rho[\gamma_{\vec k}^\mu, \gamma^{\nu}_{-\vec k}])
\end{equation}
where $\gamma^\mu(\vec k) = \frac{1}{\sqrt{L_x L_y}} \sum_{\vec x}e^{-i\vec k \cdot \vec x} \gamma^\mu(\vec x)$ with $\mu = 1$ or $2$.  
We will adopt anti-periodic boundary condition along the $x$ direction, and periodic boundary condition along the $y$ direction to avoid parity-invariant $k$-points in the study of GfTNS, following previous work~\cite{TN_Fermi_Surface}.  
Thus, $\vec k = (k_x, k_y)$ with $k_x = \frac{2\pi(n_x+\frac{1}{2})}{L_x}$ and $k_y = \frac{2\pi n_y}{L_y}$, for $n_i = 1,2,\cdots, L_i$ with $L_i$ being the number of sites along direction $i = x$ or $y$.

\subsection{Gaussian fermionic tensor network states}

We first review the construction of a generic \emph{fermionic} TNS before specializing to Gaussian states.
To construct a fermion TNS~\cite{fTNS}, one first prepares $n_v$ Majorana Bell pairs over each bond of the lattice, which make up the virtual state:   
\begin{equation}
  \rho_{\virtual} = \prod_{b\in \text{bonds}}\prod_{j=1}^{n_v} \frac{1}{2} (1+i\gamma^j_{b_1} \gamma^j_{b_2})
  \label{eq:rho_virtual}
\end{equation}
where $\gamma_{b_1}^j$ and $\gamma_{b_2}^j$ are respectively the $j$-th Majorana modes on the two sites connected by bond $b$.
Then, at each site $\vec x$, which hosts $n_p$ Majorana fermion modes (representing the physical degrees of freedom. {$n_p = 2$ for the case of one complex fermion per site.}), one prepares a state $\rho_\psi$ in the tensor product of the virtual Hilbert space and the physical Hilbert space on $\vec x$. 
Due to the assumed translation invariance, $\rho_\psi$ is the same for all $\vec x$. 
A fermion TNS supported in the physical Hilbert space, with bond dimension $\chi=\sqrt{2}^{n_v}$ \footnote{When $n_v$ is even, a GfTNS can be constructed as a conventional TNS with bond dimension $2^{n_v/2}$. When $n_v$ is odd, this is no longer possible, yet the odd-$n_v$ GfTNS is still well-defined. Thus, one should take $\chi=\sqrt{2}^{n_v}$ as a continuation from even $n_v$ to odd $n_v$.}, is formed as the partial overlap between $\rho_\virtual$ and $\prod_{\vec x} \rho_\psi$:
\begin{equation}
  \rho_{\TNS} = \Tr_{\virtual} (\rho_\virtual \prod_{\vec x} \rho_\psi)  ,
\end{equation}
where $\rho_\virtual$ is always Gaussian.

The TNS is a GfTNS when $\rho_\psi$ is also Gaussian~\cite{fTNS, GfTNS_chiral, GfTNS_dwave},~i.e.,~one whose physical properties is entirely determined, via Wick's theorem, by a covariance matrix of the form of Eq.~\ref{eq:covar}. 
From now on, we assume $\rho_\psi$ to be Gaussian and denote its covariance matrix as 
\begin{equation} \label{eq:standrad grouping}
  \Gamma_\psi = \MM{A}{B}{-B^T}{D}
\end{equation}
with $A$ being $n_p\times n_p$, $B$ being $n_p\times 4n_v$, and $D$ being $4n_v \times 4n_v$.  
Treating $\Gamma$ as a matrix with indices $\mu$ and $\nu$, the covariance matrix of the GfTNS can be obtained as~\cite{fTNS, GfTNS_chiral, GfTNS_dwave}  
\begin{equation}
  \Gamma_\TNS(\vec k) = A + B (D + \Gamma_\virtual(\vec k))^{-1} B^T
  \label{eq:TNS_virtual}
\end{equation}
where $\Gamma_\virtual(\vec k)$ is the momentum-space covariance matrix of $\rho_\virtual$ and is equal to 
\begin{equation}
    \Gamma_\text{virtual}(\vec k) = \MM{0}{\Gamma_1(\vec k)}{-\Gamma_1(\vec k)^\dag}{0}, \hspace{5mm} 
\end{equation} 
where $\Gamma_1(\vec k) = -(e^{-ik_x} I_{2n_v} \oplus e^{-ik_y}I_{2n_v})$ with $I_{2n_v}$ being the $2n_v \times 2n_v$ identity matrix.
{In writing $\Gamma_\text{virtual}(\vec k)$, we have ordered the virtual modes in the order of $l, d, r, u$ legs shown in Fig. \ref{fig:isoT}.}

\subsection{Isometric constraints on GfTNS}
\label{subsec:isometric_contrains}
\begin{figure}[tb]
    \hspace{-10mm}
  \begin{subfigure}{0.20\textwidth}
  \includegraphics[scale=0.8]{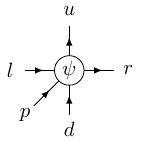}
  \end{subfigure}%
  \hspace{-5mm}
  \begin{subfigure}{0.30\textwidth}
  \includegraphics[scale=1.2]{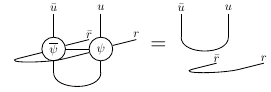}
  \end{subfigure}
\caption{Left: The isometric tensor that makes up the TNS. 
  Right: 
  The contraction of the incoming arrows of $\psi$ and $\overline{\psi}$ leaves an identity matrix on each outgoing index. 
  In equation, this is $\sum_{pld} \psi_{plrdu} \overline{\psi}_{pl\bar{r}d\bar{u}} = \delta_{r\bar{r}}\delta_{u\bar{u}}$.
  }
\label{fig:isoT}
\end{figure}

As proved in~\cite{isofTNS}, the isometric structure and the fermionic structure is compatible in a TNS, in the sense that if the fermionic TNS is made of isometric tensors, the norm of the many-body fermion wavefunction can still be computed efficiently despite of the fermionic structure. 
For our purpose, we will simply consider a local tensor that is isometric as in Fig.~\ref{fig:isoT}; this is the same isometric constraint obeyed by the black tensors in Fig.~\ref{fig:uni-isotns}.
In the context of GfTNS, on each leg of $\psi$ lives a number of Majorana fermion modes.  
Fig. \ref{fig:isoT} simply means that the reduced density matrix of the modes on the right and the up legs is proportional to the identity, and thus that the covariance matrix entries vanish; $\frac{i}{2}\Tr(\rho_\psi [\gamma^\mu, \gamma^\nu]) \propto \frac{i}{2}\Tr([\gamma^\mu, \gamma^\nu]) = 0$ for all $\gamma^\mu$ and $\gamma^\nu$ on the right and up legs.  

The reverse statement is also true: if $\rho_\psi$ is Gaussian, and the block of the covariance matrix according to certain tensor legs is zero, then the tensor must be isometric with those legs having outgoing isometric arrows.
This follows from the Wick's theorem: all higher-order correlators among the outgoing legs in $\psi$ are equal to the Pfaffian of the corresponding submatrix of the covariance matrix of the outgoing legs, which is identically zero. 
The full set of correlators determines the reduced density matrix entirely, which is the identity matrix, implying isometry.   

We order the Majorana fermions in the order of the legs of $\psi$: $p, l, d, r, u$, and use $n_p, n_l, n_d, n_r, n_u$ to denote the number of Majorana modes on each leg.
According to the discussion above, the covariance matrix of an isometric tensor will have a zero block if we regroup the blocks according to incoming and outgoing arrows.
The covariance matrix of an isometric $\rho_\psi$ would have the additional block structure:  
\begin{equation}
  \Gamma_{\psi} = \MM{A}{B}{-B^T}{D} = \MM{\A}{\B}{-\B^T}{\D}  = \MM{\A}{\B}{-\B^T}{0} .
  \label{eq:G_block}
\end{equation}
The $A,B,D$ follow the grouping separating physical and virtual fermions as in Eq.~\eqref{eq:standrad grouping}, where $A$ is of size $n_p \times n_p$ and $B$ is of size $n_p \times (n_l+n_d+n_r+n_u)$.
The $\A,\B,\D$ follow the grouping separating incoming and outgoing indices, where $\A$ is of size $n_\inn \times n_\inn$, $\B$ is of size $n_\inn \times n_\out$ with incoming dimension $n_\inn = n_p + n_l + n_d$ and outgoing dimension $n_\out = n_u + n_r$. 
The $\D$ is sub-block of $D$ that vanishes due to isometric constraint.

To construct an uniform isoGfTNS as in Fig. \ref{fig:uni-isotns}, one simply uses a covariance matrix whose sub-block corresponding to the right and up virtual legs are zero. 
To construct an alternating isoGfTNS, one uses a unit-cell of two tensors (see Appendix \ref{subsec:unit_cell}), with different outgoing isometric arrows as in Fig. \ref{fig:alt-isotns}, and the corresponding sub-block of the covariance matrices of $\psi$ vanishing: for one site in the unit-cell, the sub-block of the right and up virtual legs is zero; for the other site, the sub-block of the right and down virtual legs is zero. 

\subsubsection{Variational parametrization}
In the following, we show how to parametrize $\Gamma_\psi$ for isoGfTNS.
Note that there are two other constraints on $\Gamma_\psi$ in addition to the isometric constraints.
The first is that $\Gamma_\psi \Gamma_\psi^T = I$ due to that $\rho_\psi$ is a pure state.
The second is that $\Gamma_\psi$ is real antisymmetric~\cite{Bravyi}. 
The antisymmetry requirement simply means that $\A$ is real antisymmetric. 
The purity requirement is equivalent to 
\begin{enumerate}
  \renewcommand{\labelenumi}{\Roman{enumi}.}
  \item $\B^T\B = I$
  \item $\A^T\A = I - \B\B^T$ 
  \item $\A \B = 0$
\end{enumerate}

Condition I means that $\B$ is an isometric matrix.
To unpack condition II, using the antisymmetry of $\A$, let us write (assuming $n_\inn$ is even) 
\begin{equation}
  \A = R \bigoplus_{i=1}^{n_\inn/2} \MM{0}{s_i}{-s_i}{0} R^T, \hspace{5mm} s_i \ge 0
\end{equation}
where $R$ is orthogonal. 
Then condition II becomes
\begin{equation}
  \A^T \A = R \bigoplus_{i=1}  \MM{s_i^2}{0}{0}{s_i^2} R^T = I - \B\B^T = I - \text{Proj}_\B
\end{equation} 
which is an orthogonal projection matrix whose rank is $n_\inn - \rank(\B) = n_\inn - n_\out$.  
It follows that $s_i = 1$ for $i \le \rank(\A)/2$, and $s_i = 0$ for $i > \rank(\A)/2$, and that the column space of $R$ spans the orthogonal subspace to the column space of $\B$. 
Condition III follows directly from II. 

Thus, to parametrize $\Gamma_\psi$, we consider an orthogonal matrix $Q$ of size $n_\inn \times n_\inn$ and take the first $n_\out$ columns to be $\B$ and the other $n_\inn - n_\out$ columns to be $R$: 
\begin{equation}
  Q = [\hspace{1mm}\B \hspace{1mm} |\hspace{1mm} R\hspace{1mm}], \sp \A = R \bigoplus_{i=1}^{n_{\text{in}/2}} \MM{0}{1}{-1}{0} R^T
  \label{eq:Q}
\end{equation}

In the end, we note that if there is an even number of sites in the lattice and $Q$ is the same on all sites, one can always choose $Q$ to be a special orthogonal matrix,~i.e.,~$\det(Q) = 1$, for all tensors in the TNS. 
Suppose a link in the lattice connects sites $a$ and $b$. 
The GfTNS state does not change if one swaps the order of two modes on both sites. 
This permutes the columns and the rows of $\Gamma_a$ and $\Gamma_b$ accordingly, which results in a permutation only on the columns of $Q$, flipping the determinant of $Q$. 
The only invariant in such operations is the parity of the sum of the determinants of all the $Q$s.  
Following this, the determinant of the $Q$ on every site can be converted to 1, if one has an even number of sites in the system. 
In the following, we will always use an even number of sites and use only special orthogonal $Q$s. 

\subsubsection{Gradient-based energy minimization}
To find the ground state within the isometric GfTNS manifold, one directly works with $Q$ and does Riemannian minimization of the total energy.   
Given the system Hamiltonian 
\begin{equation}
  H = \sum_{\vec k} \sum_{\mu\nu}\gamma^{\mu\dag}(\vec k) h(\vec k)_{\mu\nu} \gamma^\nu(\vec k), 
\end{equation}
the expectation value of $H$ in a Gaussian state $\Gamma_\TNS$ is~\cite{TN_Fermi_Surface} 
\begin{equation}
  \braket{H} = \sum_{k} i\Tr[h(\vec k) \Gamma_\TNS(\vec k)] + \Tr(h(\vec k)) 
\end{equation}
where $\Gamma_\TNS$ is a function of $Q$ via Eq. \ref{eq:TNS_virtual}, \ref{eq:G_block}, and \ref{eq:Q}.  
In the following we will use isoGfTNS to study the representability of isoTNS.  
See Sec. \ref{sec:optimization} for details of the numerical optimization.

\subsection{Gaussian constraints on isoTNS}
Here we comment that Gaussian constraints restrict isoTNS more than general TNS. 
This can be seen from a counting argument on the dimension of the TNS and isoTNS variational manifold. 
The dimension of the variational manifold of all TNS of certain bond dimension $\chi$ is given by the dimension of the manifold of the TNS tensors minus the dimension of the manifold of the gauge freedom on the virtual bond of the TNS. 
For example, for a translationally invariant MPS of bond dimension $d\times \chi \times \chi$, with $d$ being the dimension of the local Hilbert space, the MPS tensors form a $d\chi^2$-dimensional complex manifold.    
The gauge freedom on the virtual bonds are the $\chi \times \chi$ invertible matrices, which, as a manifold, has complex dimension $\chi^2$. 
Thus, 
\begin{equation}
  \dim_\mathbb{C} \text{MPS} = (d-1) \chi^2 
\end{equation}
For isometric MPS, the tensors form a $(d\chi,\chi)$-Stiefel manifold, which has effective complex dimension $d\chi^2 - \chi^2/2$. 
The gauge freedom are the $(\chi,\chi)$ unitary matrices, which has effective complex dimension $\chi^2/2$. 
Thus, 
\begin{equation}
  \dim_\mathbb{C} \text{isometric MPS} = (d-1) \chi^2 = \dim_\mathbb{C} \text{MPS} 
\end{equation}
which makes sense because every MPS can be brought into canonical form without increasing the bond dimension. 

For 2D general TNS and isoTNS, similarly, the dimensions are respectively 
\begin{equation}
  \dim_\mathbb{C} \text{TNS} = d\chi^4 - 2 \chi^2 
\end{equation}
\begin{equation}
  \dim_\mathbb{C} \text{isoTNS} = \left(d-\frac{1}{2}\right) \chi^4 - \chi^2
\end{equation}
They clearly differ, but the difference is quite mild. 
To acheive the same variational dimension, when $\chi$ is large, the bond dimension would scale as  
\begin{equation}
  \chi_{\text{isoTNS}} = \left(\frac{d}{d-\frac{1}{2}}\right)^{1/4} \chi_\text{TNS} \underset{d=2}{\approx} 1.07 \chi_\text{TNS} 
  \label{eq:1.07}
\end{equation}

For Gaussian TNS with $2$ physical Majorana modes and $n$ virtual modes on each leg, the variational  manifolds for general GfTNS and isoGfTNS are respectively $\text{SO}(4n+2)$ and $\text{SO}(2n+2)$, giving
\begin{equation}
  \dim_\mathbb{R} \text{Gaussian TNS} = \frac{(4n+2)(4n+1)}{2} 
\end{equation}
\begin{equation}
  \dim_\mathbb{R} \text{Gaussian isoTNS} = \frac{(2n+2)(2n+1)}{2} 
\end{equation}
Asympototically for large $n$, if Gaussian TNS and Gaussian isoTNS are to have the same variational dimension, then 
\begin{equation}
  n_\text{isoTNS} = 2 n_\text{TNS} \rightarrow \chi_\text{isoTNS} = \chi_\text{TNS}^2
\end{equation}
which is a much worse scaling than Eq. \ref{eq:1.07}. 
One thus expects that the Gaussian constraint {restricts isoTNS more than regular TNS.}

\section{Numerical Results}
\label{sec:results}
Using the framework and algorithms of non-interacting isoGfTNS and interacting isoTNS, we compare the performance of the uniform and alternating ansatzes on various paradigmatic systems, including (i) the stacked chains example from Sec.~\ref{sec:diagonal}, (ii) the Fermi surface, (iii) the band insulator, (iv) the $p_x+ip_y$ mean-field superconductor, and (v) the transverse field Ising model. 
In each case, we find that alt-iso(Gf)TNS outperforms uni-iso(Gf)TNS, providing evidence that alt-isoTNS has more representative power than uni-isoTNS.
{Additionally for the free fermionic systems, we compare to GfTNS, which has no isometric constraint, to demonstrate the loss in representability from imposing the isometric constraints necessary for quantum circuit equivalence.}

\subsection{Stacked Chains \label{sec:free_fermion_hopping_1_1}}
\begin{figure}[t]
  \hspace*{-0.75cm}
  \includegraphics[scale=1.]{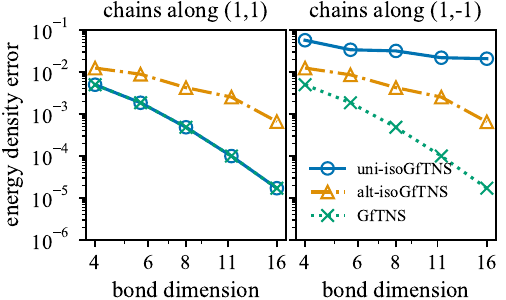}
\caption{Error in ground state energy for a $36\times36$ free fermion system with hopping along (1,1) or (1,-1) diagonals, using GfTNS. While isoTNS shows strong dependence on the entanglement structure of the state, alt-isoTNS is insensitive to this.
}
\label{fig:FS_diagonal}
\end{figure}

To test our argument of the entanglement structure in Sec.~\ref{sec:diagonal}, we consider a collection of 1D free fermion states hopping only along the (1,1)-diagonals or the (1,-1)-diagonals on a 2D square lattice. 
Using the isoGfTNS technique, the ground state is computed as both an uni-isoTNS and an alt-isoTNS.
The energy result is presented in Fig.~\ref{fig:FS_diagonal}. 
We also plotted the generic {GfTNS} result without any isometric constraints. 
We see when the entanglement is only along the (1,1)-diagonals, the uni-isoTNS is actually as accurate as the unconstrained TNS and better than the alt-isoTNS.
But when the entanglement is along the (1,-1) direction, uni-isoTNS performs poorly while alt-isoTNS better approximates the exact results.
Also, the performance of alt-isoTNS is the same for both (1,1) and (1,-1) chains, indicating that the ansatz is insenstive to the directionality of entanglement.
This strongly supports our arguments laid out in Sec.~\ref{sec:diagonal}.

In practice for generic Hamiltonians, the states one encounters will have entanglement along all directions, and thus we expect the alt-isoTNS to be a better ansatz than the uni-isoTNS. We demonstrate this below for paradigmatic examples of non-interacting and interacting systems.

\subsection{Fermi Surface}
Consider a two-dimensional square lattice of free fermions with Hamiltonian: 
\begin{equation}
  H = -\sum_{\braket{i,j}} c_i^\dag c_j
  \label{eq:fermi_surface}
\end{equation}
where $c_i^\dag$ and $c_j$ are fermion creation and annihilation operators, and $\braket{i,j}$ denotes a pair of nearest neighbor sites. 
This is a system of free fermions whose ground state has a Fermi surface. 
Systems with a Fermi surface are among the most challenging to represent as a tensor network state, because they are very strongly entangled; the entanglement entropy of a subregion with a boundary of size $A$ scales as $A\log(A)$~\cite{FS_EE_I, FS_EE_II}, exceeding the conventional area-law for ground states.
We use this system as a benchmark to see the difference in the ability of uni-isoTNS and alt-isoTNS to represent strongly-entangled systems.  

\begin{figure}[ht]
\centering
\includegraphics[scale=0.95]{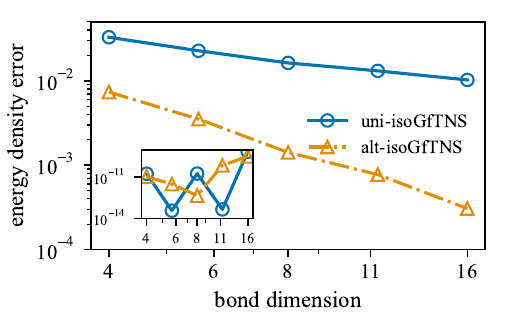}
\caption{Error in ground state energy for a $96\times96$ Fermi surface on a square lattice, see Eq. ~\ref{eq:fermi_surface}. Inset: norm of the energy gradient as a function of bond dimension{, indicating that our isoGfTNS numerics are well converged.}}
\label{fig:FS}
\end{figure}

We follow the setup in~\cite{TN_Fermi_Surface} which studied the representability of generic, non-isometric TNS for Fermi surface using GfTNS. 
As in~\cite{TN_Fermi_Surface}, we also study a system with an anti-periodic boundary in $x$ condition and periodic boundary in $y$ direction, {which results in an open shell structure for the Fermi surface.}
We work directly in the momentum space.   
The isoTNS with this boundary condition does not have an {OHS}, but we expect that the representability for large 2D systems is mainly limited by the isometries in the bulk. 
The calculation is performed for bond dimensions $\chi = \sqrt{2}^n, n = 4,5,..,8,$ for both uni-isoTNS and alt-isoTNS, of size $96\times 96$, with isometric arrows drawn in Fig. \ref{fig:isotns}(ab) (excluding the OHS,~i.e.,~red tensors); {as before, $n$ is the number of pairs of Majorana Bell pairs used to represent each bond, which corresponds to $\sfrac{n}{2}$ complex fermion Bell pairs.}
As shown in Fig.~\ref{fig:isotns}(c) and Fig.~\ref{fig:FS}, the alt-isoTNS captures the Fermi surface profoundly better than the uni-isoTNS. 
In particular, for uni-isoTNS, there is an apparent directionality in the sharpness of the Fermi surface \footnote{From basic solid state physics, the directionality in the $k$-space translates into the same directionality in the real space}.
The entanglement perpendicular to the direction of the isometric arrows is represented much worse, as explained by the decoupled chain argument in Sec.~\ref{sec:diagonal}.
{The inset of Fig. \ref{fig:FS} shows the norm of the energy gradient as a function of the bond dimension, which is a measure of how well converged our gradient descent optimized isoGfTNS results are. The combination of initializing the gradient descent with 30 to 40 initial Gaussian states, the global nature of the gradient updates, and the small gradient norms together are an indication that we likely have found the true global minima.}

\subsection{Gapped free fermion systems}
\begin{figure}[htb!]
  \begin{subfigure}{0.5\textwidth}
  \includegraphics[scale=0.95]{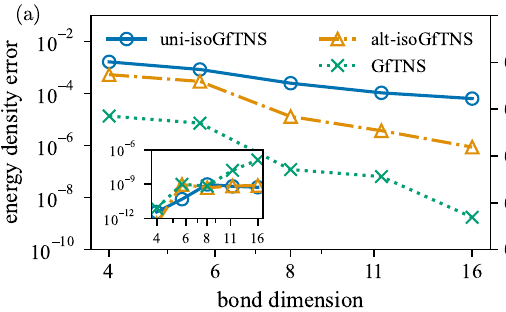}
  \end{subfigure}
  \begin{subfigure}{0.5\textwidth}
  \includegraphics[scale=0.95]{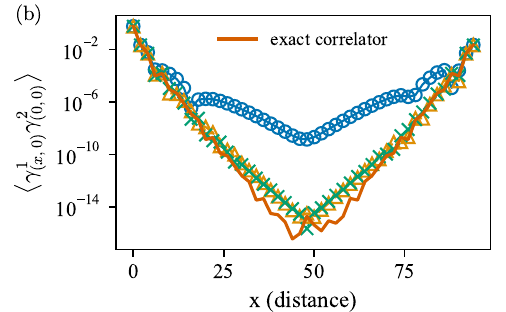}
  \end{subfigure}
\caption{Result of the GfTNS calculation for a $96\times96$ free fermion insulator.
(a) Error in ground state energy density. Inset: norm of the energy gradient as a function of bond dimension. 
(b) Correlator
  $\braket{\gamma^1_{(x,0)} \gamma^2_{(0,0)}}$ along the $x$-axis, as a function of the $x$-distance. 
  The bond dimension is 16.
  }
\label{fig:insulator}
\end{figure}

We present additional evidence for the better representability of alt-isoTNS for less entangled, gapped free fermion systems.  
We study a topologically trivial band insulator with a finite band gap with the Hamiltonian 
\begin{equation}
  H_{\text{Insulator}} = - \sum_{\braket{i,j}} c_i^\dag c_j + \sum_{i}\mu_i c_i^\dag c_i  
\end{equation}
with $\mu_i = 1$ on the even sublattice, and $-1$ on the odd sublattice. 
The ground state and two-point correlator on a 96 $\times$ 96 lattice, with periodic boundary condition along $x$ and anti-periodic boundary condition along $y$, are computed with GfTNS, and the results are shown in Fig. \ref{fig:insulator}. 
As seen, alt-isoTNS also better approximates both properties of the insulator than uni-isoTNS does with the same bond dimension.
Although both alt- and uni-isoTNS have a larger energy density error than the unconstrained TNS, we see in Fig. \ref{fig:insulator}(b) that alt-isoTNS accurately captures long-distance two-point correlations in the state, far better than uni-isoTNS does.

\subsection{$p_x + ip_y$ mean-field superconductor}
\begin{figure}[htb!]
\includegraphics[width=0.45\textwidth]{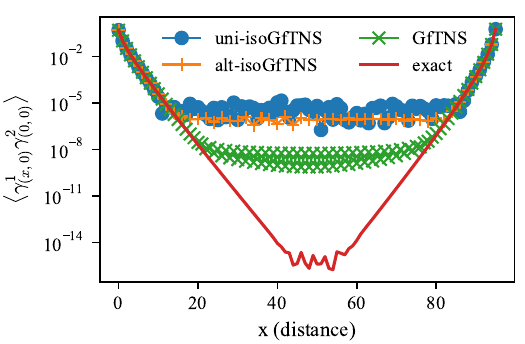}
  \caption{
  Correlator
  $\braket{\gamma^1_{(x,0)} \gamma^2_{(0,0)}}$ along the $x$-axis, as a function of the $x$-distance on a $96\times96$ free fermion $p_x+ip_y$ mean-field superconductor. Bond dimension is 16.  
  {The $\sim 10^{-15}$ error of the exact correlator is due to the floating-point error of the double-precesion we adopted in our numerical calculation.}
}
\label{fig:PipSC}
\end{figure}

The mean-field $p_x+ip_y$ superconductor (PipSC) has the following chiral Hamiltonian:
\begin{equation}
  H_{\text{PipSC}} = - \sum_{\braket{i,j}} c_i^\dag c_j + 2 \sum_{i} c_i^\dag c_i + \sum_{\braket{i,j}} (D_{ij} c_i c_j + h.c.),
\end{equation}
with $D_{ij} = 1$ on the horizontal bonds and $\sqrt{-1}$ on the vertical bonds.
If placed on a space manifold with no boundary, it has a finite energy gap and thus exponentially decaying correlators.  
PipSC, however, is topologically non-trivial,~i.e.,~not in the same phase as a product state.
It has been shown that 2D GfTNS cannot be gapped, topologically non-trivial, {\it and} have a local parent Hamiltonian at the same time~\cite{GfTNS_chiral, Dubail_Read}.  
Here, we investigate the performance of GfTNS in approximating the ground state of this chiral Hamiltonian.
The ground state on a 96 $\times$ 96 lattice is computed with GfTNS, and the results are shown in Fig. \ref{fig:PipSC}. 
As we see, there is a clear length scale for all three kinds of GfTNS beyond which the correlator stops decaying expoentially. 
This resembles the behavior of recent numerical studies of 2D TNS gapped chiral states using variational infinite projected entangled pair states (PEPS), where the correlators decay exponentially for short distances and then eventually algebraically~\cite{chiral_iPEPS_I, chiral_iPEPS_II}.  

\begin{figure}[htb]
  \begin{subfigure}[T]{0.25\textwidth}
  \includegraphics[scale=0.8]{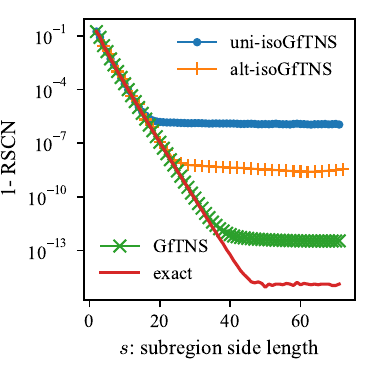}
  \end{subfigure}%
  \begin{subfigure}[T]{0.25\textwidth}
  \includegraphics[scale=0.7]{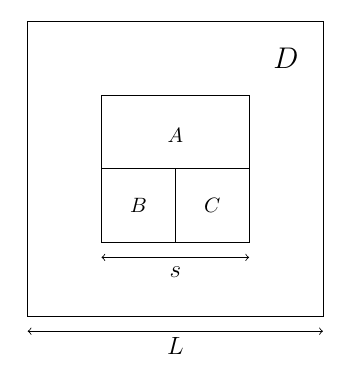}
  \end{subfigure}
\caption{
Left: The real-space Chern number computed as a function of the subregion size. 
Right: Partition of the subregions.
The GfTNS has bond dimension 16, and is on a $96 \times 96$ lattice. 
}
\label{fig:RSCN}
\end{figure}

To access the ``chiralness'' of the TNS, one can compute the real-space Chern number (RSCN)~\cite{Kitaev_anyon} of the optimized GfTNS.
To compute RSCN, one partitions the spatial region of a pure state into four regions $A, B, C, D$ such that $A, B, C$ form a triple joint and $D$ complements $ABC$; see the configuration in Fig. \ref{fig:RSCN}.  
The RSCN in the Majorana basis is defined as 
\begin{equation}
  \nu = \frac{1}{16} 12 \pi i [\Tr(PP_APP_BPP_C) - \Tr(PP_CPP_BPP_A)]
\end{equation}
where $P_{ij} = \braket{\gamma_j\gamma_i}$ is the Majorana correlator matrix. 
$P_A,P_B,P_C$ are the spatial projectors onto the corresponding regions,~i.e.,~diagonal matrices with $i$-th diagonal elements equal to 1 if site-$i$ is in the region, to 0 if not. 
Kitaev proved that if the correlators decays exponentially, then as the size of all subregions goes to infinity, $\nu$ approaches an integer, and is proportional to the Hall conductance of the system~\cite{Kitaev_anyon}. 
As seen, the various types of GfTNS approximate well the exact Chern number of PipSC, which equals 1, but the error to the exact value saturates to a finite non-zero, albeit small, value.
{While the advantge of alt-isoTNS is less pronounced in Fig. \ref{fig:PipSC}, we see that alt-isoTNS captures the chiralness of the state significantly better in Fig. \ref{fig:RSCN}.}

It is, however, unlikely that the numerical optimization of a GfTNS can find a true gapless topologically non-trivial state, because such states are always non-injective~\cite{GfTNS_chiral}, which have measure 0 in the variational manifold. 
Thus, the length scale in both the correlator and the RSCN is a result of the tension between good quantitative approximation of the chiral ground state and the general inability of the GfTNS to represent a gapped chiral state.  
The tensor network representation of chiral states deserves further study.

\subsection{DMRG\textsuperscript{2} study of 2D critical transverse field Ising model \label{sec:DMRG}}

\begin{figure}[h]
\includegraphics[width=0.48\textwidth]{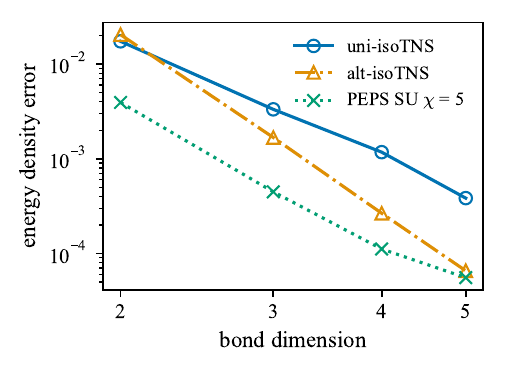}
\begin{center}
\caption{
DMRG\textsuperscript{2} results for critical 2D transverse field Ising model ($g = 3.04438$) on a $12\times12$ square lattice with open boundary condition. The accuracy of alt-isoTNS increases with bond dimension faster than that of uni-isoTNS. {For the ground truth energies, we use 1D DMRG on a $\chi=2048$ MPS, which agrees within statistical error to quantum Monte Carlo results.
}
PEPS simple update results at the same bond dimensions are provided for comparison.
}
\label{FigDMRGresult}
\end{center}
\end{figure}

To compare the representing power of uni-isoTNS and alt-isoTNS for interacting spin systems, we use the $\text{DMRG}^2$ algorithm~\cite{lin2021efficient} to find the ground states using these two ansatzes.
The $\text{DMRG}^2$ algorithm is essentially a nested loop over two levels of DMRG. 
We optimize each column sequentially, sweeping left to right and vice-versa. 
Within each column, we find the ground state of the effective Hamiltonian of that column using traditional 1D DMRG.
At a high level, $\text{DMRG}^2$ is analogous to the standard DMRG sweeping algorithm if we view each column as a site, now with each site being updated itself with DMRG.
As we do the left-right sweep, we use MM to shift the orthogonality column to the column we are optimizing. The isometric form provides stability to the $\text{DMRG}^2$ algorithm as in the 1D case.
The algorithm is essentially the same for alt-isoTNS and uni-isoTNS, with the small change of MM mentioned in Sec.~\ref{subsec:alt-isotns-algorithm}.

As a concrete example, we study the 2D transverse field Ising (TFI) model at criticality on a square lattice with open boundary conditions,
\begin{equation}
    H = -\sum_{\langle ij \rangle} Z_i Z_j + g\sum_{i}X_i
\end{equation}
where $g = 3.04438$.
We chose the system size to be $12\times12$ so that 1D snake DMRG with {$\chi=2048$} gives accurate ground state energy, {within statistical error of $10^{-5}$} with quantum Monte Carlo {with $3\times 10^{9}$ cluster updates}.
For the 2D isoTNS, we set the bond dimension to be $\chi$ everywhere except on the orthogonality column (row), where we use a larger bond dimension $\chi^2$.
As shown in Fig.~\ref{FigDMRGresult}, for $\chi = 2, 3, 4, 5$, the energy density converges as a power-law of the bond dimension. Interestingly, the power for alt-isoTNS is considerably larger than the power for uni-isoTNS.
{Moreover, for uni-isoTNS, as we change the horizontal isometric arrows from leftward to rightward during the sweep, we need to change the vertical isometric arrows from upward to downward. Otherwise, uni-isoTNS $\text{DMRG}^2$ does not converge. See Sec.\ref{app:DMRGsweep} for details of isometric directions during the DMRG sweep.}
This result suggests that alt-isoTNS does have an algorithmic advantage in finding the ground state of local isotropic Hamiltonians.
{
While at this system size we do not see an advantange over 1D DMRG, we expect one to develop as $L_y \rightarrow \infty$.
Using imaginary time TEBD\text{2} to optimize ground states of the critical TFI, an infinite strip uni-isoTNS finds a lower energy (when compared to QMC) than infinite 1D DMRG when $L_y=20$~\cite{iMM}.
}

To place the accuracy of the isoTNS results in context, we also present the PEPS simple update (SU) \cite{simple_update} results for comparison \footnote{The PEPS SU data are provided by an anonymous referee, for whom we are grateful.}.
As seen in Fig. \ref{FigDMRGresult}, at bond dimension $5$, alt-isoTNS is comparable to PEPS SU, although slightly worse.  
We expect the more costly PEPS full update~\cite{algorithm_finite_peps} to give more accurate results than SU. 
However, unconstrained PEPS still lack the ability to be efficiently prepared on a quantum computer, which remains a key advantage of the isoTNS ansatz.

\section{Discussion} 
\label{sec:discussion} 

In this work, we first introduced the alternating isoTNS ansatz.
Secondly, we extended the {GfTNS} framework, a standard tool for {efficiently} investigating tensor network representability, to include isometric constraints, resulting in the isoGfTNS ansatz.
Using a pedagogical example of stacked chains, we argued that entanglement in isoTNS is mediated along the isometric arrows, and thus, unlike the uni-isoTNS, the alt-isoTNS admit efficient paths to mediate entanglement both parallel and orthogonal to the isometric constraints in the network.
As generic states are expected to have entanglement in all directions, we expect alt-isoTNS to perform better than uni-isoTNS, as we demonstrated in a number of both non-interacting and interacting examples.
Using the introduced isoGfTNS framework, we demonstrated that alt-isoGfTNS outperform uni-isoGfTNS in representing the ground states of paradigmatic free-fermion systems, namely (1) decoupled stacks of chains, (2) the Fermi surface, (3) the gapped band insulator, and (4) the $p_x+ip_y$ mean-field superconductor.
We expect a similar advantage in the representation power for {interacting} alt-isoTNS over fermionic uni-isoTNS based on the argument that the limiting factor for isoTNS is not interaction but entanglement~\cite{TN_Fermi_Surface}.
Turning to the transverse field Ising model, an interacting spin model, we showed that the energy density error of alt-isoTNS converges with a larger power law than that of uni-isoTNS.
This indicates that, despite the two ansatz having the same leading order computational cost and retaining the same benefits of optimization in an orthonormal basis as compared to generic TNS, alt-isoTNS has an representation advantage over uni-isoTNS.
Finally, from a quantum circuit perspective, we showed that alt-isoTNS can be sequentially generated as a quantum circuit with a depth of $O(L^2)$ compared to $O(L)$ of the uni-isoTNS.
{The advantage of alt-isoTNS over uni-isoTNS implies that given the same number of local gates, different circuit depths and different order of applying the gates significantly changes the result of state-preparation on a quantum computer.}

Alt-isoTNS, or more generally the non-uniform variation of isometric constraints on a TNS, opens up many interesting research directions. 
On the theoretical side, a key question is whether there are phases of matter that can be efficiently represented by alt-isoTNS but not by uni-isoTNS, or vice versa.
While uni-isoTNS were recently shown to be able to represent critical states with algebraic correlations, arising from classical partition functions viewed as quantum states~\cite{liu2023topological}, the representability of chiral states is more subtle.
For example, in a related paper to appear, some of us show that uni-isoGfTNS cannot represent \textit{gapless} chiral free fermion systems, even though general GfTNS can~\cite{GfTNS_chiral, Dubail_Read}.
In this context, an important relevant question is whether gapless chiral states admit alt-isoGfTNS representations.

On the practical side, due to the improved performance and negligible computational overhead, one should use alt-isoTNS in the place of uni-isoTNS for many-body simulations. 
A generalization that could improve further the representability is to allow the vertical isometric arrows to alternate even more freely,~i.e.,~not just alternate between each column. 
In such a ``free form" isoTNS, the optimization algorithm may automatically determine which pattern of vertical isometric arrow directions are the best for the particular system being simulated.
{For example, when sweeping across the columns left to right (and vice versa), one can choose the column arrow direction based on the Moses Move error for the two-column splitting, i.e. always choose the direction giving the smaller error.}
Another promising direction is the optimization of isoTNS with variational Monte Carlo (VMC). 
Due to its low computational complexity in bond dimension, variational Monte Carlo for tensor network states has recently become the state of the art in simulating a TNS ground state in two dimension~\cite{PEPS_VMC}.
A key step in VMC algorithms is the sampling of basis states according to their Born probabilities. All uni-isoTNS, alt-isoTNS, and ``free form" isoTNS can be more efficiently sampled than generic TNS~\cite{Vieijra_2021, anand2024METTS}, therefore it is natural to extend VMC to isoTNS where again alt-isoTNS is expected to provide an improvement to uni-isoTNS.

The variety of isometric constraints on TNS can go beyond uniform, alternating, and free forms.
Going one step further,
one may consider networks composed of tensors that are isometric upon several different groupings of legs.
One example is given in
the recently proposed dual-isometric projected entangled pair states~\cite{yu2024dualisometric}, which allow for efficient calculation of a broader class of observables than previously possible with conventional isoTNS.
The isoGfTNS approach we have introduced here offers two distinct advantages: 
\begin{enumerate}
    \item 
    In general, each of the TNS with different isometric constraints requires a specific optimization algorithm utilizing the isometric condition.
    {For example,} the algorithm for dual-isometric projected entangled pair states~\cite{yu2024dualisometric} is distinct to that for uni-isoTNS.
    However, this is not the case in the framework of GfTNS.
    One does not need to develop different optimization algorithms for different isometric constraints.
    GfTNS with arbitrary isometric constraints can all be parameterized by simply demanding that sub-blocks of the correlation matrix corresponding to outgoing legs are zero, as discussed in Sec.~\ref{subsec:isometric_contrains}, and optimized accordingly.
    This enables high throughput evaluation of many different isometric constraints. 

    \item In general, algorithms for TNS with different isometric constraints, such as those discussed in Sec.~\ref{subsec:alt-isotns-algorithm} and gradient descent algorithms for dual-isometric projected entangled pair states, have a complexity polynomial in bond dimension.
    Our isoGfTNS algorithms have a huge advantage in that complexity scaling is polylogarithmic in bond dimension.

\end{enumerate}
Together, this establishes isoGfTNS as a tool of independent interest in studying the representability of TNS ansatzes with various isometric constraints.

Although in this work we focused on 2D systems on square lattices, a further interesting future direction is to investigate the effect of isometric constraints on TNS representation in higher dimension or different lattice geometry.
The complexity of GfTNS and isoGfTNS scales much more favorably, only polynomially with dimension, making such a study possible.

Finally, due to the strong connection between isoTNS and unitary circuits and the ability to efficiently optimize isoTNS on classical computers, the study of various isometric networks will shed light on quantum state preparation on near-term quantum computers.
In addition, the alt-isoTNS ansatz finds potential applications in machine learning, particularly for scenarios where models are trained classically but inference is performed on quantum devices.
Algorithms for manipulating classical isoTNS can be interpreted as approximation algorithms for quantum circuit compilation.
For example, one can imagine the approximation algorithm using MMs to transform an alt-isoTNS to an uni-isoTNS corresponds to a quantum circuit compilation of approximately transforming an $O(L^2)$ depth circuit into $O(L)$ depth.
Looking forward, further exploration of isoTNS may provide valuable insights into scalable quantum state preparation and the development of quantum-assisted machine learning applications.

\section{Acknowledgement}
Y.W. is supported by a start-up grant from IOP-CAS.
L.W. and Q. Y. are supported by the National Natural Science Foundation of China under Grants No. T2225018, No. 92270107, and No. 12188101, No. T2121001, and the Strategic Priority Research Program of Chinese Academy of Sciences under Grants No. XDB0500000 and No. XDB30000000. 
ZD, SA and MZ were supported by the U.S. Department of Energy, Office of Science, Basic Energy Sciences, under Early Career Award No. DE-SC0022716.
S.L. and F.P. were supported by the DFG TRR80. 
F.P acknowledge support from the Deutsche Forschungsgemeinschaft (DFG, German Research Foundation) under Germany’s Excellence Strategy--EXC--2111--390814868, TRR 360 – 492547816 and DFG Research Unit FOR 5522 (project-id 499180199), the European Research Council (ERC) under the European Union’s Horizon 2020 research and innovation programme (grant agreement No. 771537), as well as the Munich Quantum Valley, which is supported by the Bavarian state government with funds from the Hightech Agenda Bayern Plus.
Y.W. is grateful for discussion with Hong-Hao Tu. 
The authors are grateful for an anonymous referee for providing the PEPS simple update data for comparison with the isoTNS interacting results. 
The data that support the findings of this article are openly available \cite{data}; embargo periods may apply.
\section{Appendix}
\subsection{GfTNS with unit cell}
\label{subsec:unit_cell}
When there are unit cells in the system, let $\gamma^\mu_{\vec r}$ denote a Majorana fermion mode where $\vec r$ denotes the unit-cell position, and $\mu$ collectively denotes all other intra-cell degrees of freedom such as the position within a unit-cell and the first or the second part of a complex fermion mode. 
On each site $\vec x$, one still assigns a local Gaussian linear map with covariant matrix $A(\vec x)$, and take 
\begin{equation}
A \equiv A(\vec r) = \bigoplus_{\vec x:\text{in unit cell $\vec r$}} A(\vec x)
\end{equation}
where translational invariance demands $A(\vec r)$ be the same for all $\vec r$. 
Let $B$ and $D$ be similarly defined. 
Then the physical covariance matrix of the TNS is still given by Eq. \ref{eq:TNS_virtual}.

\subsection{Larger MERA for $\ket{\negdiag}$}
\begin{figure*}[h]
  \hspace*{-0.75cm}
  \includegraphics[scale=1.]{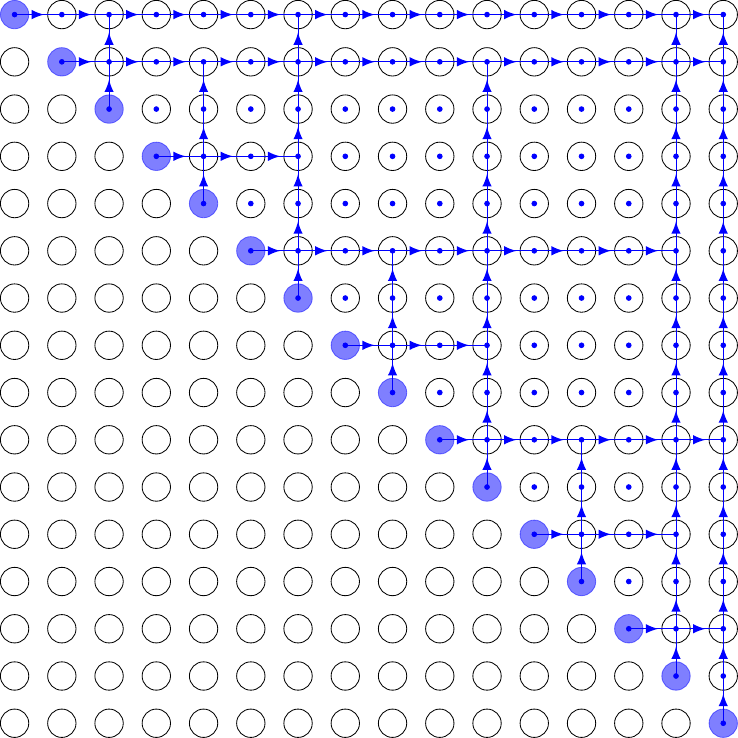}
\caption{MERA representing a single $(1,-1)$ diagonal chain, embdeded on a $16 \times 16$ square lattice.
}
\label{fig:large_mera}
\end{figure*}

As discussed in Section~\ref{sec:diagonal}, we can understand the representational difference between alt- and uni-isoTNS by considering stacks of diagonal chains. 
For the negative diagonal state $\ket{\negdiag}$, uni-isoTNS most efficiently represents this state as a MERA.
In Fig.~\ref{fig:large_mera}, we demonstrate this for a single chain of length $L=16$. 
For this larger system size than $L=5$ shown in Fig.~\ref{fig:uni-mps-negdiag}, we see the alternating layers of unitaries and isometries.
If we were to stack the MERA for all possible $(1,-1)$ diagonal chains, we would see that the tensor in the top-right corner, the {OC}, would require virtual bond dimensions of $\chi^{L-1}$, where $\chi$ is the bond dimension of each MERA.

{
\subsection{Isometric directions during a $\text{DMRG}^2$ sweep}
\label{app:DMRGsweep}

\begin{figure*}[htb]
  \begin{subfigure}{0.5\textwidth}
  \includegraphics[scale=0.38]{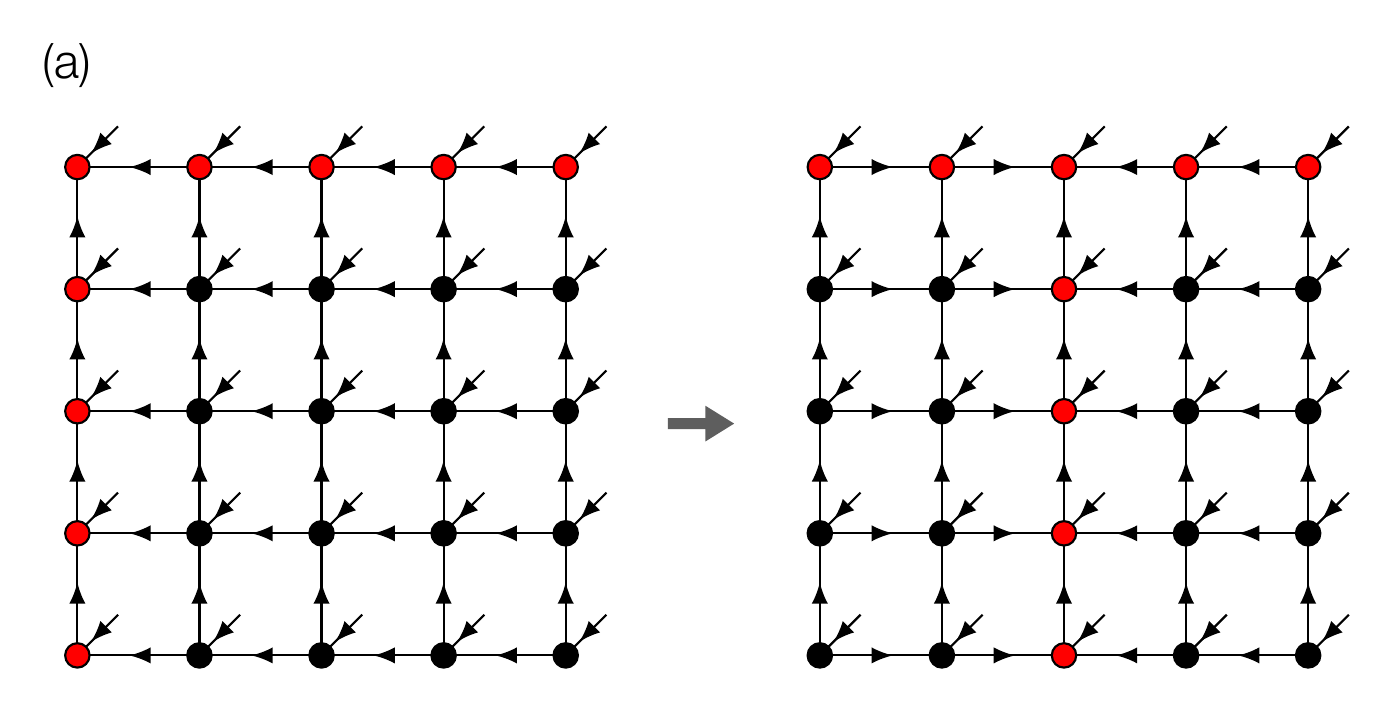}
  \end{subfigure}%
  \begin{subfigure}{0.5\textwidth}
  \includegraphics[scale=0.38]{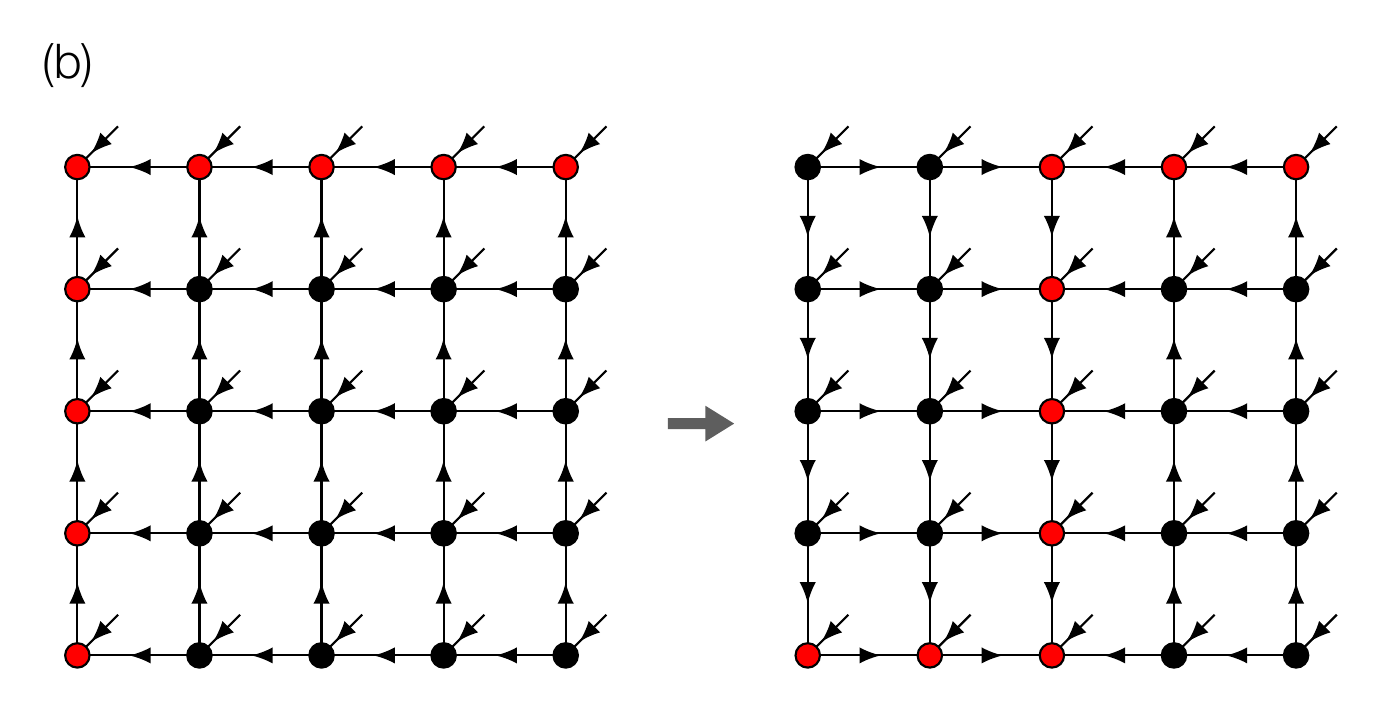}
  \end{subfigure}
  \begin{subfigure}{0.5\textwidth}
  \includegraphics[scale=0.38]{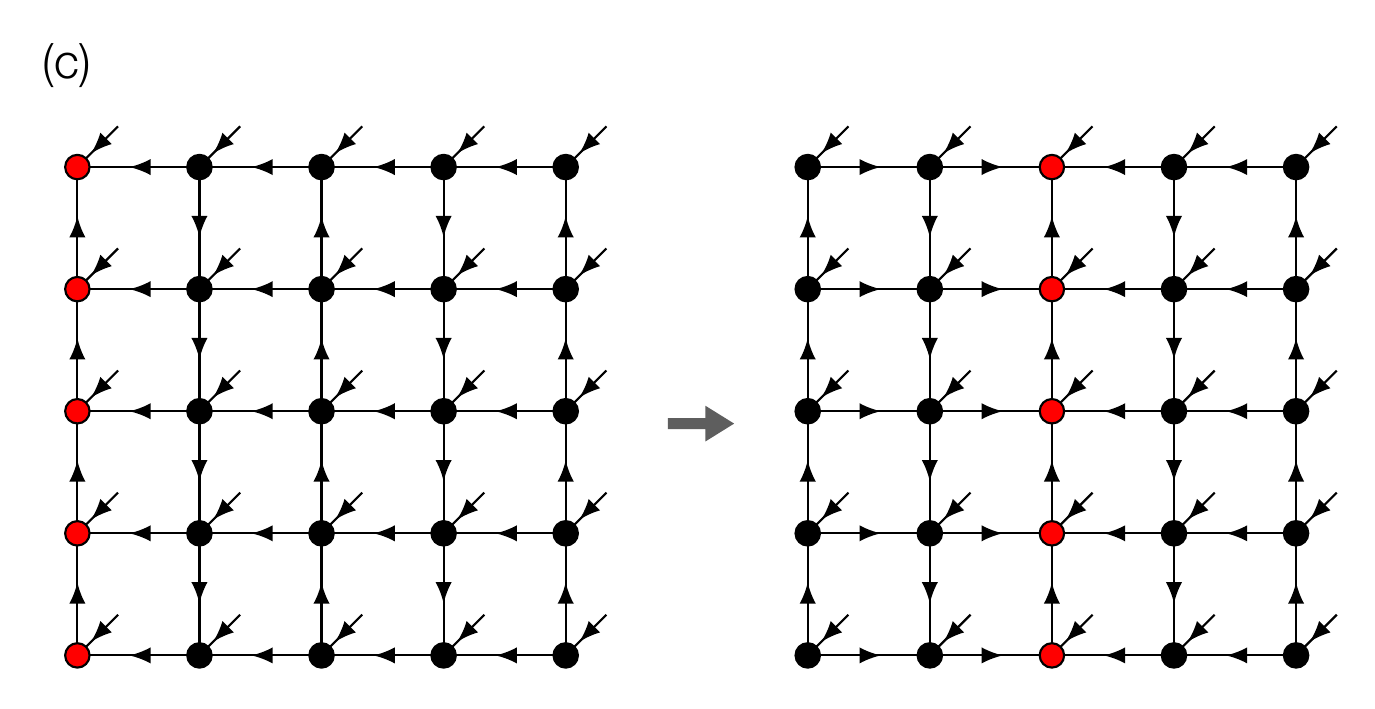}
  \end{subfigure}%
  \begin{subfigure}{0.5\textwidth}
  \includegraphics[scale=0.38]{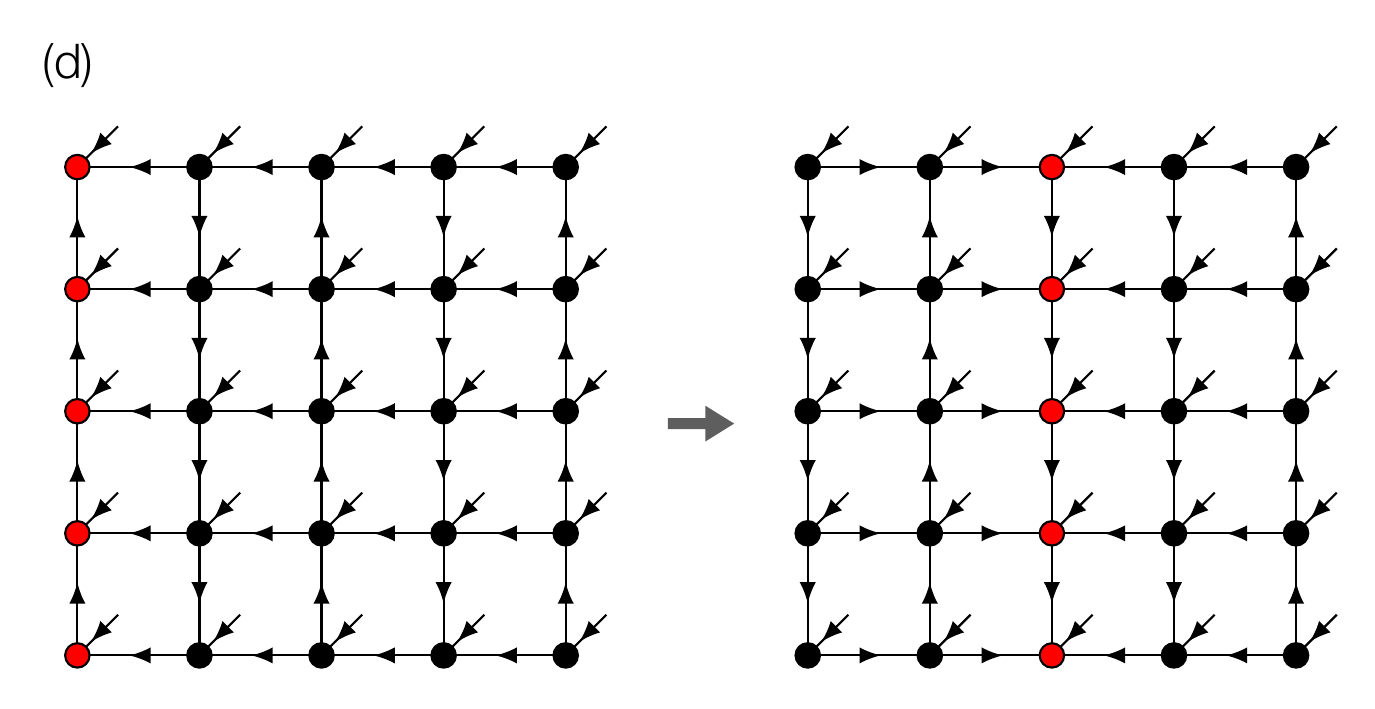}
  \end{subfigure}
\caption{isometric directions as we move the orthogonality column from the left to the middle during a \DMRGsq sweep (a) uni-isoTNS, $\uparrow$ to $\uparrow$ sweep (b) uni-isoTNS, $\uparrow$ to $\downarrow$ sweep (c) alt-isoTNS, $\uparrow\downarrow$ to $\uparrow\downarrow$ sweep (d) alt-isoTNS, $\uparrow\downarrow$ to $\downarrow\uparrow$ sweep}
\label{fig:DMRGsweepisometricarrows}
\end{figure*}

\begin{figure}[h]
\includegraphics[width=0.5\textwidth]{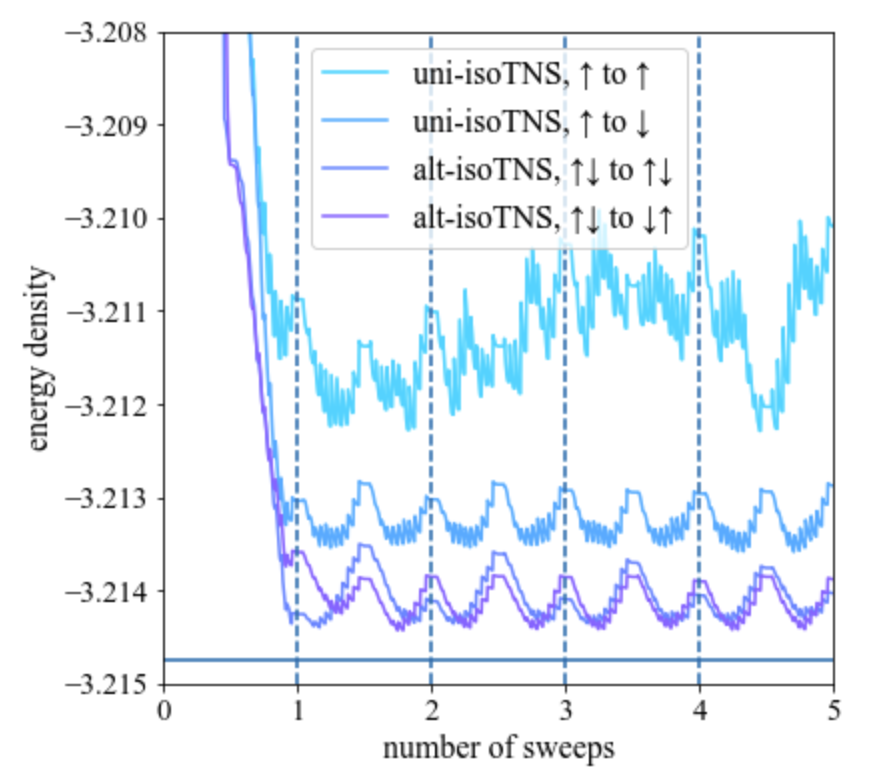}
\begin{center}
\caption{Energy density during \DMRGsq sweeps of the critical TFI model on a 12 $\times$ 12 square lattice. Five full sweeps are shown in the figure. In each full sweep, the orthogonality column is moved from the left to the right and back to the left. Different colors indicate different vertical isometric directions during the sweep. The bond dimension of isoTNS is 4, and the bond dimension along the orthogonality column is 16.}
\label{fig:DMRGsweepdependence}
\end{center}
\end{figure}

In this appendix, we discuss different choices of isometric directions during a \DMRGsq sweep and how they change the result. As illustrated in Fig.~\ref{fig:DMRGsweepisometricarrows}, as we move the orthogonality column and optimize the tensors along the column, the horizontal arrows are required to point into the orthogonality column, but we can choose the direction of the vertical arrows. For uni-isoTNS with isometry arrows originally pointing up and to the left, we can either keep the vertical arrows up during the sweep, namely an $\uparrow$ to $\uparrow$ sweep (Fig.~\ref{fig:DMRGsweepisometricarrows}(a)), or reverse the vertical arrows as we sweep, an $\uparrow$ to $\downarrow$ sweep (Fig.~\ref{fig:DMRGsweepisometricarrows}(b)).
Similarly, for alt-isoTNS, we can choose between the $\uparrow\downarrow$ to $\uparrow\downarrow$ sweep (Fig.~\ref{fig:DMRGsweepisometricarrows}(c)) and the $\uparrow\downarrow$ to $\downarrow\uparrow$ sweep (Fig.~\ref{fig:DMRGsweepisometricarrows}(d)).

Since we have shown that the isometric direction affects the representing power, it is interesting to test how they change the result of \DMRGsq. 
In Fig.~\ref{fig:DMRGsweepdependence} we show the \DMRGsq result for the 4 different choices of isometric directions.
Here, we plot the energy as a function of the sweep index.
In each sweep, the orthogonality column is moved from left to the right and back again, so a non-integer value is when the OHS is away from the left edge of the system.
The $\uparrow$ to $\downarrow$ sweep for uni-isoTNS and the two types of alt-isoTNS sweeps converge to stable oscillations, but the $\uparrow$ to $\uparrow$ sweep for uni-isoTNS does not converge. This is most likely because the time direction, along which entanglement is mediated, constantly changes during the $\uparrow$ to $\uparrow$ sweep, while other choices do not suffer from this problem.
The stable oscillation patterns indicate that the position of the orthogonality column also affects the representing power of the isoTNS.
Finally, note that the energy is often minimized when the OHS in in the middle of the system (near sweep indices $n+\sfrac{1}{4}$ and $n+\sfrac{3}{4}$ for integer $n$ as the OHS will be in the middle of the 2D system), as this minimizes the average distance most tensors are from the OHS.
Finally, note that the energy is often minimized when the OHS in in the middle of the system (near sweep indices $n+\sfrac{1}{4}$ and $n+\sfrac{3}{4}$ for integer $n$ as the OHS will be in the middle of the 2D system), as this minimizes the average distance most tensors are from the OHS.
}

{
\subsection{Optimization details for GfTNS}
\label{sec:optimization}
We also comment how the GfTNS numerical optimization is done in practice. 
The optimization is non-convex.
To overcome this, one can seed the simulation with different initial states.
In our case, 30 to 40 different random initial states were each independently optimized for 10000 iterations, and then the one with the lowest energy was picked to be further optimized for 100000 iterations.
Every 10000 iterations, we also add some randomness to the state to escape local minima.
Then after that, if gradient is still large, we use Newton's method to do a final descent. 
After these steps, the energy gradients of the optimized states can often go below $10^{-9}$. 
The optimization is done on a GPU, and the computation is parallelized over the different $\vec k$ points of the translationally invariant systems.
}

%

\end{document}